\documentclass[fleqn,usenatbib]{mnras}

\usepackage{newtxtext,newtxmath}
\usepackage[T1]{fontenc}
\usepackage{ae,aecompl}

\usepackage{graphicx}	% Including figure files
\usepackage{amsmath}	% Advanced maths commands
\usepackage{amssymb}	% Extra maths symbols
\usepackage{array}
\usepackage{multirow}

\usepackage[normalem]{ulem}
\usepackage{xcolor}
\newcommand{\Dnu}{\Delta\nu}
\newcommand{\dnu}{\delta\nu}
\newcommand{\Mo}{\rm{M}_\odot}

\newcommand{\Teff}{T_{\rm{eff}}}
\newcommand{\Fe}{\left[\rm{Fe/H} \right]}
\newcommand{\nmx}{\nu_{\rm{max}}}
\newcolumntype{P}[1]{>{\raggedright\arraybackslash}p{#1}}
\newcolumntype{C}[1]{>{\centering\arraybackslash}m{#1}}
\newcommand{\SPI}{{\small{SPI}} }
\newcommand{\unidam}{{\textsc{UniDAM}} }
\newcommand{\TESS}{{\emph{TESS}} }
\newcommand{\Kepler}{{\emph{Kepler}}}
\newcommand{\corot}{{\emph{CoRoT}}}

\title[CBM with p-modes]{Convective boundary mixing in low- and intermediate-mass stars I. \\ Core properties from pressure-mode asteroseismology}  
\author[G. C. Angelou et al.]{George C. Angelou,$^{1, 3}$\thanks{E-mail: gangelou@mpa-garching.mpg.de}
Earl P. Bellinger$^{2}$,
Saskia Hekker$^{3, 2}$, Alexey Mints$^{3, 4}$,
\newauthor
 Yvonne Elsworth$^{5, 2}$, Sarbani Basu$^6$ and
Achim Weiss$^{1}$ \\
$^{1}$Max-Planck Institut f\"{u}r Astrophysik, Karl-Schwarzschild-Str. 1, 85741 Garching, Germany \\
$^{2}$Stellar Astrophysics Centre, Department of Physics and Astronomy, Aarhus University,\\ \quad Ny Munkegade 120, DK-8000 Aarhus C, Denmark\\
$^{3}$Max-Planck-Institut f\"{u}r Sonnensystemforschung, Justus-von-Liebig-Weg 3, 37077 G\"{o}ttingen, Germany \\
$^{4}$Leibniz Institute for Astrophyics Potsdam (AIP), Ander Sternwarte 16, D-14482 Potsdam, Germany \\
$^{5}$School of Physics and Astronomy, University of Birmingham, Birmingham B15 2TT, UK \\
$^{6}$Department of Astronomy, Yale University, New Haven, Connecticut, USA \\
}

\date{Accepted XXX. Received YYY; in original form ZZZ}

\pubyear{2019}

\begin{document}
\label{firstpage}
\pagerange{\pageref{firstpage}--\pageref{lastpage}}
\maketitle

\begin{abstract}
% Convective boundary mixing (CBM) in stellar models is necessary in order to match observable constraints from clusters and binaries as well as individual stars. 
Convective boundary mixing (CBM) is ubiquitous in stellar evolution.
It is a necessary ingredient in the models in order to match observational constraints from clusters, binaries and single stars alike.  
We compute `effective overshoot' measures that reflect the extent of mixing and which can differ significantly from the input overshoot values set in the stellar evolution codes.
We use constraints from pressure modes to infer the CBM properties of \Kepler \ and \corot \ main-sequence and subgiant oscillators, as well as in two radial velocity targets (Procyon A and $\alpha$~Cen~A). 
Collectively these targets allow us to identify how measurement precision, stellar spectral type, and overshoot implementation impact the asteroseismic solution. 
With these new measures we find that the `effective overshoot' for most stars is in line with physical expectations and calibrations from binaries and clusters. 
However, two F-stars in the \corot \ field (HD~49933 and HD~181906) still necessitate high overshoot in the models.
Due to short mode lifetimes, mode identification can be difficult in these stars. 
We demonstrate that an incongruence between the radial and non-radial modes drives the asteroseismic solution to extreme structures with highly-efficient CBM as an inevitable outcome.
Understanding the cause of seemingly anomalous physics for such stars is vital for inferring accurate stellar parameters from  \TESS data with comparable timeseries length. 
\end{abstract}


\section{Introduction}
Convective boundary mixing (CBM) remains a major uncertainty in stellar modelling.  
A range of behaviours is possible when convective elements (e.g., plumes, eddies) encounter stable boundaries \citep{1991A&A...252..179Z, 2015A&A...580A..61V}. 
The outcomes depend on factors including, but not limited to, the momenta of the convective material, the underlying fluid properties and the 
 stratifications in the adjacent radiative region. 
These behaviours are described by the non-local and time-dependent Navier-Stokes equations for hydrodynamics. 
However, it is challenging to capture the complexity of these equations in a 1D formalism as is required by current stellar evolution calculations \citep{1997ApJ...489L..71C}. 
Instead, most common convective theories are local and time-independent, thus necessitating an explicit formalism for describing CBM.
Understanding, parameterizing and calibrating CBM currently drives much work in stellar physics and hydrodynamics alike. 
This is due in part to the fact that the inclusion of CBM in stellar models is necessary to match observations of stars across different mass ranges and phases of evolution (see \S \ref{sec:obsev}).

%However, there are cases when asteroseismic fitting can result in CBM efficiencies far higher than those inferred from other well-constrained systems such as binaries or clusters (see below).
%Here we investigate several factors that dictate the asteroseismic solution.  
%The goal of this study is to reliably and accurately determine the impact of CBM through asteroseismic modelling.
%The goal of this study is to reliably and accurately determine the impact of CBM through asteroseismic modelling. 

Asteroseismology is able to shed light on these uncertain CBM processes by providing constraints on the internal stellar structure.
In this work, we modify our chosen stellar evolution code (MESA, \citealt{2018ApJS..234...34P}) and asteroseismic pipeline (Stellar Parameters in an Instant, SPI, \citealt{2016ApJ...830...31B}) to better diagnose the CBM regions in asteroseismic targets. As is common in stellar evolution, CBM in our models is parameterized with an overshoot formalism. 
As we discuss below, the implementation and efficiency of overshoot can vary from code to code much like convection theory in general. 
The values from one code are not necessarily transferable, other than to say the overshoot efficiency needed to match an observation is high, low, or consistent with zero. 
In our analysis we provide several measures of overshoot and the fully-mixed region in order to make our results more comparable in the literature. 

We are motivated by the fact that asteroseismic fitting can result in CBM efficiencies far higher than those inferred from well-tried calibration methods such as fitting to binaries or clusters. 
Typically an extension of the mixed region above the convective core of $0.1-0.3 H_P$ is required in these systems \citep{2015A&A...575A.117S, 2017ApJ...841...69R} whereas, in some cases, factors two to five times larger have been found from asteroseismic inference (see for example \citealt{2014ApJ...780..152L, 2014ApJ...787..164G} and our results below). There is also contention as to whether the efficiency of CBM exhibits a linear dependency with mass (at $M < 2.0 \Mo$), with several calibration studies arriving at opposing conclusions  \citep{2015A&A...575A.117S,2018A&A...618A.177C, 2018ApJ...859..100C}. 
In this work we rely on pressure modes (p-modes) to infer CBM properties which limits our scope to lower-mass targets ($M < 1.8$M$_{\odot}$).

We analyse \Kepler, \corot \ and radial-velocity (RV) main-sequence and subgiant stars. These targets form a modest sample, useful in that they differ in their combination of observable features and stellar types. 
The stars are well-studied having been modelled by different groups, thus providing a rich body of literature with which to compare various methodologies and assumptions. 
We consider the role of asteroseismic precision, stellar spectral type, and how the underlying implementation of overshoot impacts upon our ability to constrain their CBM parameters -- the results of which have important implications for the spate of \TESS data currently being released.

%Our manuscript is organized such that in \S\ref{sec:obsev} we outline the importance of CBM for stellar evolution in general. In \S \ref{sec:CBMOS} we summarise how overshoot is implemented in stellar evolution codes. 
%In \S \ref{sec:Numerics} we describe the numerical codes used in this work and their modifications for the current study. 
%In \S \ref{sec:Targets} we outline the targets of analysis and their different data sources. We report our findings in \S \ref{sec:results} and provide a brief discussion in \S \ref{sec:discussion} before summarizing in \S \ref{sec:conclusions}. 
%A detailed comparison and validation of our pipeline is provided in the supplementary online material. 

\section{Observational evidence for CBM in stellar evolution} 
\label{sec:obsev}

\subsection{Modification to the thermal stratification through CBM}

In stellar evolution, CBM is traditionally parameterized as overshoot by extending the convection zone beyond the stable border. 
It is commonly also implemented as a weak diffusive process by either increasing the size of the fully-mixed region or assuming the mixing velocity of the overshooting material decays exponentially past the boundary (see \S \ref{sec:num_os} for details). With no theory to describe the efficiency of heat penetration, one is free to treat the overshoot region as either adiabatic or radiative.   

Due to the modern tendency to treat convective mixing through a diffusion equation, generally the overshoot region is left as radiative\footnote{With operator splitting in stellar evolution the simplest approach is to modify the diffusion coefficients outside the convection zone.}. 
It is hence implicitly assumed that the convective efficiency is low enough that there is no feedback on thermal structure. 
%In stellar evolution, CBM is most often parameterized as overshoot by assuming that the convective efficiency is low enough that there is no feedback on thermal structure. 
However, in reality, such an assumption is only appropriate in the low P\'{e}clet\footnote{The  P\'{e}clet number is the comparison of the thermal diffusion timescale to the convective turnover timescale.  It indicates the impact of radiative effects on the material and thus gives insight on the behaviour of  material entering a stably stratified region. } number (P\'{e}) regime \citep{1991A&A...252..179Z, 2015A&A...580A..61V, 2017LRCA....3....1K}, for example, 
when overshooting at the top of the convective zone \citep{2015A&A...578A.106A} where P\'{e} is $\mathcal{O}(10)$.
Convection deeper in the stellar interior is characterized by high Reynolds numbers. By its nature it is highly turbulent, giving rise to a spectrum of eddy sizes and plume properties. 
On scales where  P\'{e} $\gg 1$, i.e., where the convective turn-over timescale is shorter than the diffusion timescale, penetrating material can modify the entropy in the stably stratified region. 
%At the base of solar convective envelope, P\'{e} is $\mathcal{O}(10^5)$ \citep{2017LRCA....3....1K}. 
%(P\'{e} is $\mathcal{O}(10^5)$ at the base of solar convective envelope, \citealt{2017LRCA....3....1K}). 
Material in high P\'{e} regimes, such as at the base of the solar convective envelope \citep[where P\'{e} is $\mathcal{O}(10^5)$,][]{2017LRCA....3....1K}, can continue to travel adiabatically with the overall effect of extending the convective region. %superadiabatic region.
As a consequence it weakens the %subadiabatic
stratification in stable zones, and thus gives rise to a thermal boundary or transition layer. 
The high  P\'{e} case is thus referred to as `penetrative convection' to distinguish it from `classical' overshoot in the low P\'{e} regime.
Various forms of these overshoot formalisms have been explored in the literature leading to different conclusions, consequences and uncertainties for stellar theory and nucleosynthesis.

Often our observational constraints are not sufficient to constrain the degree to which CBM modifies the thermal stratification. 
However, helioseismology can place constraints on the stratification imparted by penetrative overshoot at the base of the solar convective envelope through glitch analysis or by inverting the sound speed profile \citep{1994MNRAS.269.1137B,1997MNRAS.288..572B,2011MNRAS.414.1158C,2018MNRAS.481.4389J}.
%If a transition in the sound speed gradient or its derivatives, say through a change in the thermal stratification, 
%occurs over a length scale smaller than the typical wavelength of the modes, it will impart an oscillatory signal in the frequencies.
Many studies have investigated the overshoot region in the Sun and favor a sharp transition to a radiative stratification \citep{1994MNRAS.267..209B, 1994MNRAS.268..880R, 1995MNRAS.276..283C}.
%Many studies have implemented non-local MLT models with overshoot and favor a sharp transition to a radiative stratification \citep{1994MNRAS.267..209B, 1994MNRAS.268..880R, 1995MNRAS.276..283C}.
Typically these lead to small extensions of the base of the convective envelope with an upper limit of one tenth of a pressure scale height ($H_P$). 
 
Motivated by the fact that traditional overshooting models were marginally inconsistent with the seismic data, \citet{2011MNRAS.414.1158C} developed an 
overshoot model which allowed them to vary the length of the penetrative and thermal boundary layers and hence alter the steepness of the transition.
Their best fit model incorporated a smooth transition between gradients comprising a fully mixed region of $0.37H_P$ and a substantially subadiabatic layer above the Schwarzschild boundary that helps facilitate the smooth transition.
These results are consistent with parameterizations of overshoot based on plume models \citep{2018A&A...612A..21G,2004ApJ...607.1046R} but clearly differ from the previous helioseismic studies.

\subsection{Modification of the composition profile through CBM}

CBM is required in order for isochrones to fit the main-sequence turn-off in stellar clusters \citep{2003AJ....125..754W, 2017ApJ...841...69R}.
In more massive B-type stars, models with exponentially decaying overshoot better match the gravity-mode period spacing than those without  \citep{2018A&A...614A.128P, 2015A&A...580A..27M}.
Stellar models systematically fail to reproduce the magnitude of the bump in the luminosity function of globular clusters.
The typical discrepancies of approximately 0.3 magnitudes  \citep{1997MNRAS.285..593C,2002PASP..114..375S,2012ApJ...749..128A,2015MNRAS.450.2423A} can straightforwardly be reconciled though a deeper first dredge-up (i.e., overshooting during the deepest penetration of the convective envelope). 

During the core-helium burning phase, canonical stellar models are currently unable to reproduce the observed gravity-mode period spacing.
The differences are significant ($\Delta \Pi_1 \gtrapprox 50s$) and one possible solution to rectify the discrepancy is to include a prescription that maximally grows the convective core \citep{2015MNRAS.452..123C}. 
Overshoot during core-helium burning is further supported through the $R_2$ ratio, i.e., the ratio of asymptotic giant branch to horizontal branch star counts, which is proportional to the timescales of stars in each phase. 
This diagnostic is better reproduced by models with longer core-helium burning lifetimes, which overshoot helps facilitate \citep{2016MNRAS.456.3866C}. 
Parameterized template structures of white dwarfs suggest large cores are required to match measured oscillation frequencies. 
This too would point to some kind of overshoot during the helium-burning phase in order for the core to grow sufficiently large\footnote{In fact overshooting needs to be even more efficient than that predicted by the \citet{2015MNRAS.452..123C} prescription for stellar evolution to reproduce these cores.} \citep{2018Natur.554...73G}.

Overshooting during the asymptotic giant branch has significant consequences for the structure and nucleosynthesis of the star \citep{2000A&A...360..952H, 2009ApJ...696..797C, 2014PASA...31...30K}.  
Significant penetration is required in the models in order to facilitate efficient third-dredge up so to match the M/C transition luminosity (i.e., the oxygen-rich to carbon-rich transition luminosity for AGB stars) in Magellanic cloud clusters \citep{2012ApJ...746...20K}.  
%Significant penetration is required in the models in order to facilitate efficient third-dredge ($\lambda$) up so to match the M/C transition luminosity\footnote{The M/C transition luminosity is oxygen-rich to carbon-rich transition luminosity for AGB stars.} in Magellanic cloud clusters \citep{2012ApJ...746...20K}.  

In the case of classical pulsators, usually time-dependence rather than non-locality is the key aspect of the convection theory.
Time-dependent convection theories are required to describe the interaction between pulsation and convection \citep{Houdek2015} and reproduce the red edge of the classical instability strip \citep{1999ApJS..122..167B,2007MNRAS.378.1270X,2005MNRAS.360.1143D}. 
Non-locality and CBM, however, do play a crucial role in the modelling of these stars. 
\citet{2002ApJ...578..144K} demonstrated that extended convective cores are necessary in order to match the mass-luminosity of Cepheids in the Large Magellanic Cloud; canonical models are 20\% under-luminous for a given mass.
Ironically, this inclusion of CBM during the main sequence suppresses the blue loops that drive the evolution through the instability strip that give rise to the defining classical pulsations. 
Additional CBM at base of convective envelope \citep{1991A&A...244...95A,2004ASPC..310..489C} is required to once more induce the blue-ward migration. 

Precise modelling of detached double-lined eclipsing binaries also requires CBM in order to match the observed stellar parameters at concurrent ages \citep{2015A&A...575A.117S,2016A&A...589A..93D, 2017A&A...608A..62H,2018ApJ...859..100C,2018A&A...618A.177C, 2019ApJ...876..134C}.
Given their ability to break some of the degeneracies in the modelling, these systems offer perhaps our best constraints of main-sequence core overshoot. 
Crucially, high-resolution abundance measurements often complement their dynamical mass and radii determinations. 
Eclipsing binaries form the focus of Paper~II in this series. 

It should be stressed that whilst CBM can be invoked to reconcile all of the constraints listed, it is not the only possible explanation. 
Additional physics not limited to gravitational settling  \citep{2004ApJ...606..452M}, increased opacities  \citep{2018MNRAS.477.3845C} or mode trappings  \citep{2015MNRAS.452..123C} may also account for the discrepancies between modelling and observations. Nevertheless, as overshoot is a viable physical process we focus our attention on CBM in this work.

\section{Overshoot in the context of Convective Boundary Mixing}
\label{sec:CBMOS}
There exists no analytic treatment that can satisfactorily predict the flux carried by convection. The parameterization of convection thus remains one of the greatest sources of uncertainty in stellar evolution theory\footnote{And indeed the description of turbulent flows remains ``the most important unsolved problem of classical physics'' according to Feynman.}. Any theory implemented into 1D stellar evolution must ensure tractability over a nuclear timescale and across conditions that are  steeply stratified in density, pressure, temperature, and gravity. All the while also describing a process that is multi-dimensional, non-linear, non-local and  operating over length- and time-scales that vary by orders of magnitude. This is before considering how convection might then interact with other physics such as magnetism, rotation and pulsation.

To fully describe convective flows, the thermal and advective history of material must be considered up until the excess energy being carried is dissipated.
The heuristic simplifications in stellar convective theories, used to model the flux transport, introduce improper thermal stratifications in the structures. 
The resultant thermodynamic uncertainty is best illustrated through the so-called asteroseismic surface effect \citep{1984Sci...226..687B,1988Natur.336..634C} which can somewhat be avoided through averaging \citep{2016A&A...592A.159B,2018MNRAS.478.5650M,2018ApJ...869..135S}, patching \citep{2017MNRAS.472.3264J,2019MNRAS.484.5551J} or  coupling \citep{JA19} 3D convective-envelope simulations \citep{2015A&A...573A..89M,2014MNRAS.442..805T} to the interior 1D structure.  
 The main form of uncertainty in the mixing, and the focus of this study, arises from transportation across stable boundaries. 
The extent to which material `overshoots' the point of neutrality (i.e, the point of neutral buoyancy) in the models, and how the temperature gradient is modified in that region is contingent on the theory adopted or empirical calibration performed.

%The most  widely adopted convective frameworks include the mixing length theory \citep[MLT,][]{1925ZaMM....5..136P,1951ZA.....28..304B,1958ZA.....46..108B} and full spectrum of
%turbulence \citep[FST,][]{1991ApJ...370..295C, 1992ApJ...389..724C}.

The most  widely adopted convective frameworks include the mixing length theory \citep[MLT,][]{1925ZaMM....5..136P,1951ZA.....28..304B,1958ZA.....46..108B} and full spectrum of
turbulence \citep[FST,][]{1991ApJ...370..295C, 1992ApJ...389..724C}.  MLT is a single-eddy approach which averages the effect of turbulence into a single blob size.
The complicated behaviour of convection (plumes, eddies, energy dissipation, etc.) are  captured by geometrical assumptions that can vary from implementation to implementation (e.g., \citealt{2004cgps.book.....W} and \citealt{1958ZA.....46..108B}).
FST takes into account the theory of Richardson and Kolmogorov in that a range of blob sizes are responsible for mixing and that turbulence occurs on all length scales.  
As such, models with FST better reproduce the sub-photospheric pressure and density profiles in the Sun, demonstrating less of a surface effect.  
Note the surface effect can be reduced with  MLT  through  a  radially  dependent  mixing length  to  compensate  for  the  different eddy scales \citep{1997A&A...322..646S}.

In these theories the heat flux at any point in the convective region is given uniquely by the local temperature gradient, plus the local thermodynamic
variables. Furthermore, the stable convective boundary coincides with the location where the acceleration (buoyancy) is zero \citep{1906WisGo.195...41S,1947ApJ...105..305L}.
Such a naive application of the Schwarzschild or Ledoux criterion does not account for convective material arriving at the boundary with finite velocity. 
Conservation of momentum will ensure fluid will overshoot the point of neutrality and mix with material in the stably stratified region. 
The formulation of mixing beyond convective boundaries in 1D models must therefore consider:
\begin{enumerate}
    \item the functional form describing the mixed region and its extent;
    \item how and if the thermal stratification is modified,
\end{enumerate}
for which there is no reason to expect a simple expression due to the differences in the fluid properties and stratifications \citep{1991A&A...252..179Z, 2013ApJ...769....1V}. 

Although the term overshoot is typically used in the literature to describe these phenomena, there has been a recent push to replace it with the term convective boundary mixing \citep[see e.g.,][]{2013ApJ...762....8D,2019MNRAS.484.3921D} due to the complicated behaviours possible. Depending on the fluid properties convection may be plume or eddy dominated. 
Approaching plumes may turn over, penetrate or generate waves and in all likelihood undergo all three to some degree.
There is interplay between shear effects, waves and dynamical instabilities  \citep{2007ApJ...667..448M,2015A&A...580A..61V, 2017ApJ...848L...1R};  a consequence of Newton's laws and inevitably leading to mixing of the composition and possibly entropy \citep{1998ApJ...508L.103C, 2000A&A...360..952H}. 
Overshoot is applied in stellar models as a catch-all for all these physical processes, and as we have discussed, typically assumes the low  P\'{e} regime. 

%and suitable for the application we consider here, namely at the boundaries of  convective-cores in low-mass dwarfs.

\subsection{Numerical Implementation of Overshoot}

\label{sec:num_os}
Within stellar convective regions it is usually sufficient to assume the composition is instantly mixed\footnote{In \emph{most} phases the mixing timescale is much shorter than the burning timescale so the assumption is appropriate. In cases when the timescales are comparable such as for proton ingestion episodes \citep{2008A&A...490..769C} time dependence, and indeed some form of non-local information, can trivially be implemented through a diffusion equation (despite the fact convection is advective not diffusive).}. In such cases penetrative overshoot can straightforwardly be implemented by increasing the adiabatic extent of the convection zone by some fraction of a pressure (or density) scale height. 
Classical overshoot would only require that the homogeneously mixed composition is extended.

If one instead treats convection as a diffusive process, the assumption of a penetrative process would again require adopting the adiabatic temperature gradient in the overshoot region.
A diffusive step-overshoot formalism assumes the region is mixed such that
\begin{equation}
    D_{os} = D_0,
    \label{eq:Dos}
\end{equation}
where $D_{os}$ is the diffusion coefficient of the overshoot region and $D_0$ is the diffusion coefficient in the convection zone taken some distance from the convective boundary. 
In the early stages of evolution, $D_0$ is usually large enough that the overshoot region is homogeneously mixed during one timestep. 
Like the instantaneous step implementation, the extent of overshoot is assumed to be some fraction of the pressure scale height ($\alpha_{os}$) and calibrated to observations.

Alternatively, \citet{1997A&A...324L..81H} parameterized overshoot based on the simulations by \citet{1996A&A...313..497F}. 
Those simulations are based on convection in  A-stars and hot DA white dwarfs (i.e. low P\'{e} regime) and  demonstrated an exponential decay in the velocity of the overshooting material.
In analogy to the pressure scale height, $H_P$, a `velocity scale height', $H_v$, is defined such that  
\begin{equation} \label{eqn:os1}
H_v = f_{os} H_P
\end{equation} 
where $f_{os}$ is a scaling factor. The resulting equation for the diffusion coefficient is then
\begin{equation}
%D_{os} = D_0 \ e^{\frac{-2z}{H_v}} \label{eqn:os3}
D_{os} = D_0 \ \exp\left\{\frac{-2z}{H_v}\right\} \label{eqn:os3}
\end{equation}
where  \textit{z} is distance from the convective boundary.
This functional form has been assumed to apply to all convective boundaries and has been calibrated to different stellar environments and phases. 
The difference to Equation \ref{eq:Dos} is that the decay in the mixing velocities result a smoothly varying composition rather than a step in the mean-molecular weight profile.  
We refer the reader to \citet{2018A&A...614A.128P} for a recent summary of the possible formulations of chemical mixing at convective boundaries \citep{2007ASPC..378...43H, 2017ApJ...848L...1R}.

\subsection{Practical Implementation of Overshoot}

\label{sec:PracOs}
Whilst the numerical implementation of overshoot through Equations \ref{eq:Dos} \& \ref{eqn:os3} is straightforward, the efficiencies of overshoot reported by stellar codes are not necessarily transferable. 
One practical difference between the codes we wish to highlight is the determination of the point of neutrality.
Due to the non-linearities, the evolution of the star is sensitive to the location of the boundary in previous calculations. 
This means that the stability criteria used, the direction traversed through the grid to determine the boundary, or whether a search for convective neutrality applied \citep{1986ApJ...311..708L} can influence the evolution (e.g., change whether the star has third dredge-up, change the inferred age, see also \citealt{2014A&A...569A..63G}) and impact the degree of CBM required\footnote{The MESA developers have recently added the option to determine convective boundaries using their `predictive mixing' and `convective pre-mixing' algorithms. The algorithms are particularly important for more massive stars and later phases of evolution as they 
naturally incorporate semi-convective regions. They are based on similar principles advocated by \citet{1971Ap&SS..10..355C, 1986ApJ...311..708L} to name but a few. 
As predictive mixing was still under development at the time the training data was calculated it was not employed here.}. 
The location of the neutrally buoyant point will also depend on the convective theory used. 
For example, when trying to calibrate core CBM from  Reynolds-stress models  \citep{1986A&A...160..116K, 1997ApJS..108..529X, 1999ApJ...524..311C,2007MNRAS.375..388L, 2007MNRAS.375..403Y,2015ApJ...809...30A, 2016ApJ...818..146Z} 
the Schwarzschild boundary resides deeper in the interior when compared to the local theories. 
The  efficiency of overshoot required can also differ as the application of the above formulae require that a diffusion coefficient is taken at some arbitrary distance from the stable boundary. 
Modellers must also decide how to apply CBM in different evolutionary phases and at different interfaces (e.g., only on the main sequence, only at the core see Wagstaff, et al. submitted) which will also impact the evolution.

A subtle practical difference modellers must also consider is how to treat small convective zones where the radial extent is smaller than the calculated overshooting distance. 
This is particularly crucial when models settle on the zero-age main sequence, when the pp/CN reactions reach equilibrium and where the pressure scale height is large. 
An artificial convective core can develop from the numerical treatment of these reactions.  If allowed to overshoot, the cores can become large and self sustaining altering the evolution in a manner that is inconsistent with observation  \citep{1990A&A...240..262A}. To deal with small convective regions \textsc{MESA} (used in this work),  \textsc{YREC} \citep{2008Ap&SS.316...31D} and \textsc{Garstec} \citep{2008Ap&SS.316...99W} have algorithms implemented to geometrically limit the extent of overshoot.
The cuts are applied if the size of the proposed overshoot region is larger than the size of the existing convective zone.
In addition, by default MESA does not apply overshoot outside a minimum mass coordinate. 
This means the input overshoot can differ from the effective overshoot in some models along a track. 
This is a reasonable way to treat overshoot but requires careful consideration when benchmarking \SPI against results in the literature. 

The issue of how to interpret overshoot is also present when comparing ages from isochrone fitting. 
When constructing isochrones modellers must decide how far to overshoot in their evolution tracks as CBM is not applied uniformly across the Hertzsprung-Russell diagram. 
\textsc{Parsec} isochrones \citep{2017ApJ...835...77M} take the approach of varying overshoot with mass, metallicity and helium content. 
In a comparison study, \citet{2012MNRAS.427..127B} found overshooting a $0.5H_P$ with their formalism is equivalent to $0.25H_P$ with the parameterization used in the Geneva code \citep{1994A&AS..103...97M}.
 \textsc{BaSTi} isochrones \citep{2004ApJ...612..168P} reduce the efficiency of overshoot as the size of convective cores shrink and do not include the process in models with masses less than $1.1\Mo$. 
The Y$^2$ isochrones  \citep{2004ApJS..155..667D} similarly decrease the efficiency of overshoot with core size whilst applying a mass and metallicity dependent formalism.

These few examples illustrate the different yet equally valid approaches taken by modellers to treat CBM with an overshoot formalism.
In this work we modified MESA, and re-calculated our grid of models to provide detailed information about the CBM regions.
Our extracted quantities reported in \S \ref{sec:results} indicate, in a generalized way, the extent of CBM required in our models to match observations.

\section{Description and Modifications to the Numerical Codes} 
\label{sec:Numerics}
\subsection{Pipeline Overview} \label{sec:SPIov}
There are several approaches for determining stellar parameters from asteroseismic observations (scaling relations, grid searches, \emph{in situ} optimization and combinations thereof) each with their own advantages. 
In this work we utilize \SPI to characterise stellar targets and infer their CBM properties.
The \SPI strategy is a machine-learning based method whereby the algorithm emulates (learns) the stellar evolution parameter space. 
Random-forest (RF) regression is employed to discover data-driven relationships between the stellar parameters we wish to infer, and the observable properties which we measure. 
More specifically, \SPI is able to devise non-linear, non-parametric multiple regression models between parameters in its training grid.  The pipeline may be used  to determine the
forward relationships (ideal for rapid stellar evolution and population synthesis) or identify the inverse relationships (ideal for characterising stellar properties from observations) such as we do here. 
While other pipelines minimize an objective function and solve the forward equations, \SPI is unique in that it relies on inverse modelling to determine stellar parameters.

There is of course a need to calculate forward models to form a training set, however parameter evaluations are determined from the resultant inverse model. 
Validation tests have demonstrated the efficiency of this
strategy. By sampling the parameter space optimally, the algorithms can attain the same precision as forward modelling
searches with an order of magnitude fewer models and whilst exploring two extra dimensions \citep{2016ApJ...830...31B}.
SPI's advantages lie with its speed, the ability to straightforwardly and efficiently scale to higher dimensions, its insensitivity to spectroscopic systematic errors \citep{2019A&A...622A.130B} as well as the robust characterization of uncertainties.  

\subsection{MESA}
\label{sec:mesa}

\begin{figure}
    \centering
    \includegraphics[width=\columnwidth]{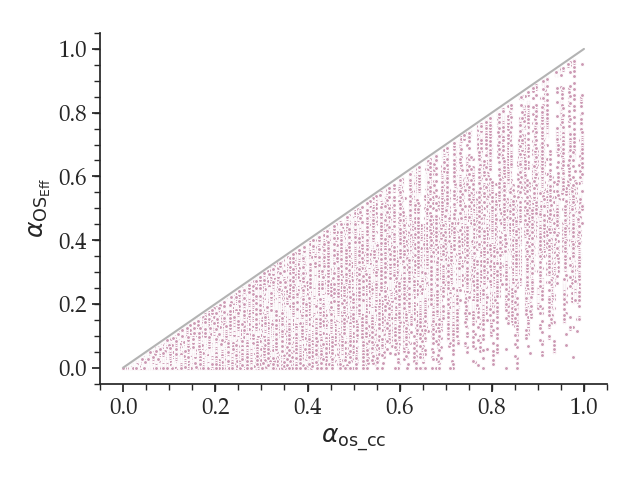}
    \caption{Differences between the input overshoot ($\alpha_{\mathrm{os\_cc}}$) and effective overshoot ($\alpha_{\rm{OS}_{\rm{Eff}}}$) for main-sequence models with convective cores. The differences are due to a combination of the $r_0$ location from which MESA overshoots and the efficiency cuts applied in the case of small convective zones. }
    %Due primarily
    \label{fig:OSIO}
\end{figure}

\begin{figure}
    \centering
        \includegraphics[width=\columnwidth]{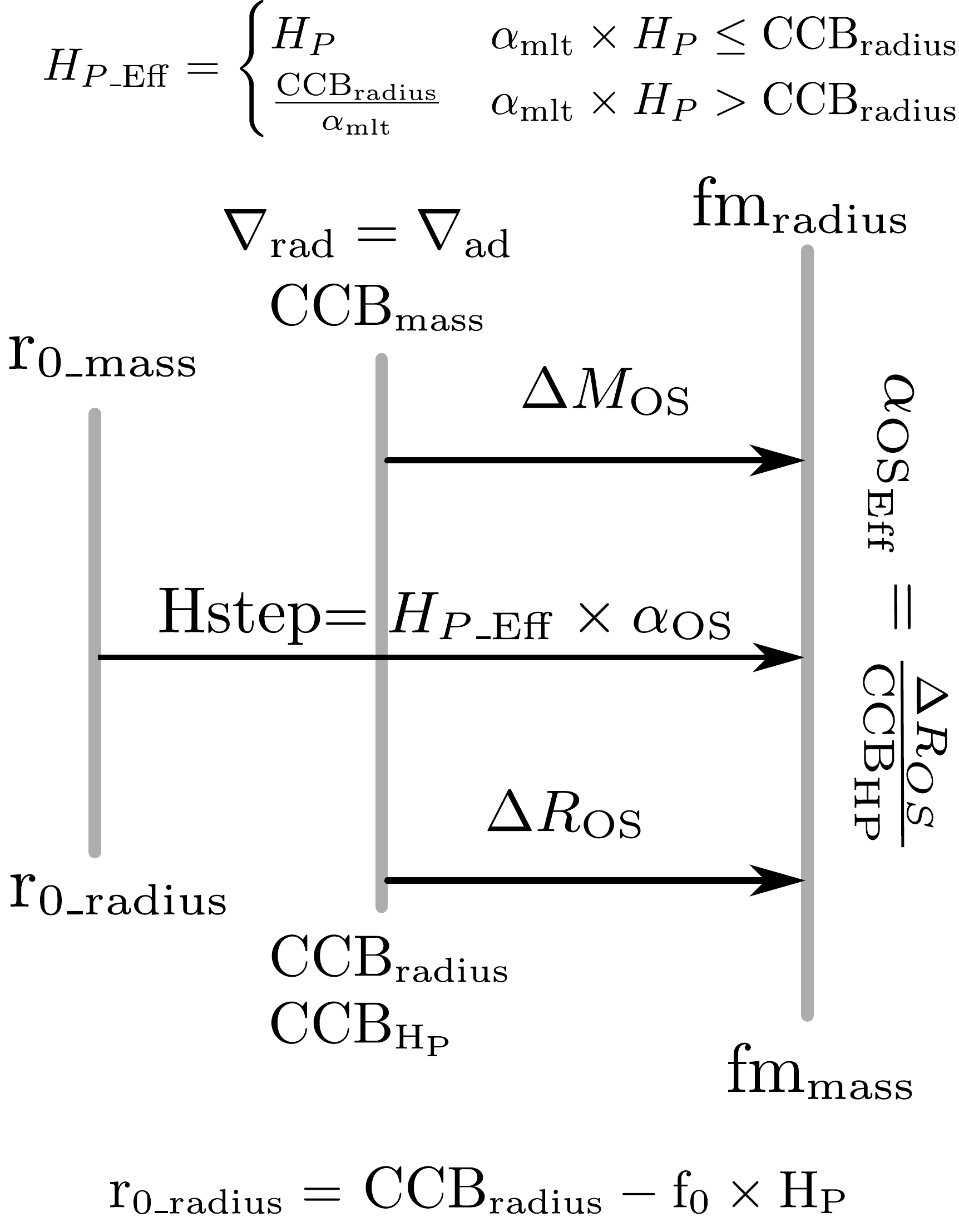}

    \caption{A schematic diagram demonstrating the quantities tracked in MESA in order to determine an effective overshoot efficiency ($\alpha_{\rm{OS}_{\rm{Eff}}}$). In addition to the usual radius and mass location of Schwarzschild boundary (CCB$_{\rm{radius}}$, CCB$_{\rm{mass}}$) we calculate the pressure scale height at that location (CCB$_{\rm{H}_{\rm{P}}}$) and the effective pressure scale height after any geometric cuts have been applied (H$_{\rm{P}\_{\rm{Eff}}}$). We follow the mass and radius location behind the stable boundary we overshoot from  (r$_{0\_ \rm{mass}}$, r$_{0\_ \rm{radius}}$), determined by stepping some fraction (f$_0$) of a pressure scale height behind the boundary. We calculate the distance over which overshoot is applied (Hstep) and the mass and radius location of the fully-mixed region (fm$_{\rm{mass}}$, fm$_{\rm{radius}}$). The mass contained in the overshoot region as well as its radial extent are also determined ($\Delta M_{\rm{OS}}$, $\Delta R_{\rm{OS}}$).} 
    \label{fig:star1}
\end{figure}

\begin{table}
    \centering
    \caption{Parameter ranges for the grid of stellar models used to train \SPI. 
        \label{tab:ranges}} 

    \begin{tabular}{lcc}%\hline\hline
        Parameter & Symbol & Range \\\hline\hline
        Mass      & $M$/M$_\odot$ & $(0.7, 1.8)$  \\
Mixing Length     & $\alpha_{\text{MLT}}$  & $(1, 3)$  \\
Initial Helium & $Y_0$  & $(0.22, 0.34)$ \\
Initial Metallicity      & $Z_0$  & $(0.0001, 0.04)^{*}$  \\
Core Overshoot                & $\alpha_{\mathrm{os\_cc}}$  & $(0.0001, 1)^{*}$ \\
Envelope Overshoot               & $\alpha_{\mathrm{os\_bce}}$  & $(0.0001, 1)^{*}$ \\
Atomic Diffusion Factor         & $D$ & $(0.0001, 3)^{*}$ \\\hline%\hline
Eff.\ Temperature        & $T_{\text{eff}}$ & $(4000, 14000)$ \\
Metallicity              & [Fe/H] & $(-2.2, 0.44)$   \\\hline

    \end{tabular}\\[0.2\baselineskip]
    \footnotesize{$^*$varied logarithmically}%\\

\end{table}

Here we build on previous studies \citep{2016ApJ...830...31B, 2017EPJWC.16005003B,2017ApJ...839..116A,2019A&A...622A.130B} to improve the scope and diagnostic potential of the \SPI pipeline.  We used \textsc{Modules for Experiments in Stellar Astrophysics} \citep[\textsc{MESA}~r11701,][]{ 2018ApJS..234...34P} to construct a training grid of stellar models as  described in \citet[][]{2016ApJ...830...31B}. 
The initial parameters were varied quasi-randomly in the ranges given in Table~\ref{tab:ranges}. 

We assume CBM is a weakly diffusive process such that $\nabla = \nabla_{\rm{rad}}$ in the overshoot region.
We apply a diffusive step-overshoot formalism (Equation \ref{eq:Dos}) at all convective boundaries.
$D_0$ is taken from a radial shell ($r_0=f_0 \cdot H_P$) inside the convective zone where it is expected that the velocity field only slightly varies \citep{1996A&A...313..497F, 1997A&A...324L..81H}.
A scaling scaling factor
\begin{equation}
 f_0=\alpha_{\mathrm{os}}/100
 \label{eqn:f0}
\end{equation} determines the distance of this point from the stable boundary. 
It is from this radial shell, $r_0$, that overshoot is applied. 
A simple example would be to set core overshoot to  $\alpha_{\mathrm{os\_cc}}=0.5$. From Equation \ref{eqn:f0} this would imply $f_0=0.005$ and the code would overshoot a distance of $0.495H_P$ from the Schwarzschild border (i.e., a distance $0.5H_P$ from a location $0.005H_P$ inside the boundary). As previously discussed, MESA may also make geometric cuts to the pressure scale height depending on the thickness of the convective zone. 
Differences between the input overshoot ($\alpha_{\mathrm{os\_cc}}$) and the effective overshoot ($\alpha_{\rm{OS}_{\rm{Eff}}}$) for main-sequence models with convective cores are shown in Figure \ref{fig:OSIO}.
\textsc{MESA's} implementation of overshoot is designed to more realistically reflect the fluid behaviour than a naive application of the 1D formalism.
As a trade off, it makes it difficult to infer an effective overshoot for comparison purposes.  
As we have noted, there are several reasons why we do not expect the `overshoot parameter' to be identical between codes. 
However, in this work we seek to report quantities that allow for a meaningful comparison and permit some degree of interpretability.

To this end we have made trivial modifications to \textsc{MESA}.
A schematic diagram of the core-overshooting region (Figure \ref{fig:star1}) illustrates some of the quantities we capture.
The most crucial of the diagnostics include the location of the Schwarzschild boundary (CCB$_{\rm{radius}}$, CCB$_{\rm{mass}}$), 
the size of the fully mixed region (fm$_{\rm{radius}}$,  fm$_{\rm{mass}}$), 
the input overshoot ($\alpha_{\mathrm{os}}$), 
the effective overshoot of the current model ($\alpha_{\rm{OS}_{\rm{Eff}}}$)  and 
the effective overshoot for the track $\left(\left<\alpha_{\rm{OS}_{\rm{Eff}}}\right>\right)$. 
The effective overshoot for the track is determined by taking the median efficiency of the constituent models that possess a convective core. 
This parameter provides perhaps a more intuitive measure of overshooting for comparison to others codes (although still approximate, recall the evolution is highly sensitive to the size of the convection zones). 
We report \SPI's accuracy in predicting these properties in the next section.

In the earlier \SPI studies overshoot was applied with equal efficiency at all convective boundaries. 
To ensure a systematic analysis of this process, we decoupled the efficiency of core-overshoot ($\alpha_{\mathrm{os}} =\alpha_{\mathrm{os\_cc}}$), and overshoot beyond the base of convective envelope ($\alpha_{\mathrm{os}}=\alpha_{\mathrm{os\_bce}}$) in the training data.
However, we do not expect to have sufficient sensitivity to the base of the convection zone.  
\SPI uses the asteroseismic ratios as input features which help to mitigate against the surface effect.
The frequency ratios, firstly, average over the glitch signal imparted by the base of the convective envelope. 
Secondly, the ratios have been shown to have stronger sensitivity to the central regions \citep{2003AA...411..215R,2005MNRAS.356..671O}.

In all, we calculated 7595 tracks evolved from the zero-age main sequence until a maximum age of $20~\text{Gyr}$ or until mixed modes were detected in the observable region  (${\nu_{\max} \pm 5\Delta\nu}$). 
From each track we extracted 32 main-sequence models (core hydrogen abundance $X_c > 0.1$) nearly evenly spaced in age, 32 turn-off models  ($0.1 \leq X_c \leq 10^{-6}$) nearly evenly spaced in $\log X_c$ and 32 sub-giant models ($X_c < 10^{-6}$) also nearly evenly spaced in age. Oscillation frequencies for these models were calculated using the GYRE code \citep[][]{2013MNRAS.435.3406T}. 
Subsequent asteroseismic quantities used throughout this manuscript are defined in Appendix \ref{sec:sdefs}. 

The extended \SPI training grid now includes new measures for CBM and comprises evolutionary models until the base of the red giant branch.
In addition we also store frequencies, ratios and mode intertias for p-modes and mixed modes (up to $\ell=3$).

\subsection{\SPI}

\subsubsection{New Input Features for the Main-Sequence}
\begin{table*}
    \centering
    \begin{tabular}{l|C{1.5cm}| P{13cm}}
        Test & No. of Features &Description of Input Features \\ \hline \hline
      T-1 &7& Global features: $\Teff$, $\Fe$, $\Dnu$, $\dnu$, $\left<r_{02}\right>$, $\left<r_{01}\right>$, $\left<r_{10}\right>$   \\ \hline
      T-2a  &13& Global features, gradients and intercepts determined from a linear fit to the local $r_{02}$, $r_{01}$, $r_{10}$ ratios as function of frequency \\  
      T-2b  &13& Same as T-2a but with ratios fit as function of radial order  \\ \hline

      T-3a &16& Global features, coefficients from a parabolic fit to the local $r_{02}$, $r_{01}$, $r_{10}$ ratios as function of frequency  \\ 
      T-3b &16& Same as T-3a but with ratios fit as function of radial order \\ \hline
      
      T-4a  &10 &Global features, coefficients from the first three principal components simultaneously fit to all $\ell=0,1,2$ modes within four radial orders of the mode closest to $\nmx$ \\
      T-4b  &10 & Same as T-4a but with modes within two radial orders of the central mode closest to $\nmx$ \\ \hline

      T-5a &10 &Global features, coefficients from the first three principal components simultaneously fit to all  $r_{02}$, $r_{01}$, $r_{10}$ ratios within four radial orders of the ratio closest to $\nmx$ \\
      T-5b  &10 & Same as  T-5a but with local ratios within two radial orders of the central ratio closest to $\nmx$  \\ \hline
    \end{tabular}
    \caption{Description of the input features considered for each test.}
    \label{tab:testDesc}
\end{table*}

As discussed in \S \ref{sec:mesa}, we now include new measures of CBM in \textsc{MESA} and our training data.
Here we construct new input features in \SPI to accurately infer these CBM parameters (Table \ref{tab:testDesc}).
The features are designed to capture the information contained in the frequencies and frequency ratios in different ways.
We devised a series of tests in order to reveal the structural information contained 
\begin{enumerate}
    \item in global asteroseismic parameters such as median separations or ratios (T-1),
    \item in local ratios through their gradients, intercepts and principal components (T-2a, 2b, 3a, 3b, 4a, 5a),
    \item in a reduced set of modes/ratios (T-4b, 5b). 
\end{enumerate}
For the time being we do not use the individual frequencies or frequency ratios as input features, as we find that for this type of methodology summary-type statistics (such as global means or medians) offer the best diagnostics.

T-2a and T-3a are motivated by \citet{2007ApJ...666..413C} who established that the gradient of the frequency differences (as a function of frequency) reflect the composition discontinuity imposed by the convective core.
 \citet{2011A&A...529A..63S} also demonstrated that the `linear regime' of the $r_{01}$ and $r_{10}$ gradients are sensitive to various core properties. 
Whilst insightful in theory, the use of (model) frequencies to infer properties about real stars will introduce a bias from the surface effect. 
Thus we also fit the gradients and intercepts as a function of radial order (T-2b and T-3b). 
In T-4a, T-4b, T-5a \& T-5b we capture the local information from the modes and ratios by taking their principal components.
The purpose of which is to reduce several dimensions of data into three summary statistics. 
For each ratio/mode we take the quantity at the radial order closest to $\nmx$.
In T-4a \& T-5a we consider the quantities four orders either side of the central value (nine modes for each of  $\ell=0,1,2$ or nine local ratios for each of r$_{01}$, r$_{10}$ and r$_{02}$).
This is truncated to two modes either side of the central value for T-4b and T-5b (five quantities for each degree or ratio).

 In each test we divided our grid into training and development (dev) sets. 
We withheld 256 dev tracks and predicted stellar parameters for the constituent models. 
In Table \ref{tab:SPIEV} we quantify how each feature contributes to the recovery of a standard set of stellar parameters.
We use the explained variance score V$_{\text{e}}$ to appraise the accuracy of the pipeline with the new CBM diagnostics included. 
The score is defined such that
\begin{equation}
  \text{V}_{\text{e}} = 1 - \frac{\text{Var}\{ y - \hat y \}}{\text{Var}\{ y \}},
\end{equation}
where $y$ is the true value we want to predict from the dev set (e.g.\ stellar mass), $\hat y$ is the predicted value from the random forest, and Var is the variance, i.e.\ the square of the standard deviation. This score indicates the extent to which the regressor has reduced the variance in the parameter it is predicting. The metric can be interpreted such that a score of one corresponds to a perfect predictor whereas V$_{\text{e}}$ $\le 0$ indicates no understanding of the parameter in question. 
\begin{table*}
\centering
\begin{tabular}{lllllllllll}
Quantity                        & Symbol                   & T-1  & T-2a  & T-2b  & T-3a  & T-3b  & T-4a  & T-4b  & T-5a  & T-5b   \\ \hline \hline
Mass                            & M$/$M$_\odot$                              & 0.961 & 0.967& 0.968& 0.966& 0.967& 0.988& 0.969& 0.966& 0.966 \\ 
Initial helium                  & Y$_0$                                      & 0.298 & 0.394& 0.413& 0.387& 0.395& 0.704& 0.422& 0.370& 0.364 \\ 
Initial metallicity             & Z$_0$                                      & 0.982 & 0.975& 0.974& 0.971& 0.969& 0.972& 0.978& 0.978& 0.978 \\ 
Mixing length                   & $\alpha_{\mathrm{MLT}}$                    & 0.363 & 0.394& 0.397& 0.394& 0.395& 0.385& 0.373& 0.380& 0.378 \\ 
Envelope overshoot              & $\alpha_{\mathrm{bce}}$                    & -0.124 & -0.096& -0.099& -0.083& -0.085& -0.122& -0.105& -0.104& -0.102 \\ 
Atomic diffusion factor         & D                                          & 0.166 & 0.209& 0.216& 0.226& 0.230& 0.188& 0.196& 0.190& 0.194 \\ 
Age                             & $\tau/$Gyr                                 & 0.972 & 0.988& 0.989& 0.990& 0.990& 0.976& 0.974& 0.985& 0.986 \\ 
Core-hydrogen abundance         & X$_{\mathrm{c}}$                           & 0.953 & 0.966& 0.968& 0.962& 0.963& 0.975& 0.961& 0.964& 0.964 \\ 
Radius                          & R$/$R$_\odot$                              & 0.995 & 0.995& 0.995& 0.994& 0.994& 0.998& 0.995& 0.995& 0.995 \\ 
Surface gravity                 & log g                                      & 0.998 & 0.998& 0.998& 0.998& 0.998& 0.999& 0.999& 0.998& 0.998 \\ 
Luminosity                      & L$/$L$_\odot$                              & 0.998 & 0.998& 0.998& 0.998& 0.998& 0.999& 0.998& 0.998& 0.998 \\ 
Surface helium                  & Y$_{\mathrm{surf}}$                        & 0.475 & 0.511& 0.525& 0.513& 0.514& 0.666& 0.531& 0.508& 0.507 \\ 
Relative radius at the BCE      & R$_{\rm{BCE}}/R_*$                         & 0.471 & 0.500& 0.500& 0.496& 0.503& 0.468& 0.473& 0.503& 0.499 \\ 
Relative mass at BCE            & M$_{\rm{BCE}}/M_*$                         & 0.456 & 0.484& 0.485& 0.480& 0.488& 0.453& 0.457& 0.488& 0.484 \\ 
Relative convective-core radius & R$_{\mathrm{cc}}/R_*$                      & 0.942 & 0.972& 0.973& 0.967& 0.967& 0.946& 0.945& 0.964& 0.962 \\ 
Relative convective-core mass   & M$_{\mathrm{cc}}/M_*$                      & 0.954 & 0.977& 0.979& 0.972& 0.973& 0.953& 0.955& 0.971& 0.971 \\ 
Core overshoot                  & $\alpha_{\mathrm{os\_cc}}$                   & 0.582 & 0.742& 0.743& 0.697& 0.704& 0.553& 0.573& 0.700& 0.703 \\ 
Effective overshoot             & $\alpha_{\rm{OS}_{\rm{Eff}}}$             & 0.655 & 0.807& 0.807& 0.786& 0.788& 0.634& 0.650& 0.767& 0.776 \\ 
Median effective overshoot      & $\left<\alpha_{\rm{OS}_{\rm{Eff}}}\right>$ & 0.635 & 0.800& 0.801& 0.769& 0.771& 0.604& 0.631& 0.764& 0.769 \\ 
Convective-core radius          & R$_{\mathrm{cc}}/R_\odot$                  & 0.965 & 0.981& 0.981& 0.976& 0.977& 0.966& 0.966& 0.978& 0.977 \\ 
Convective-core mass            & M$_{\mathrm{cc}}/\Mo$                      & 0.977 & 0.988& 0.988& 0.984& 0.984& 0.975& 0.977& 0.986& 0.985 \\ 
Fully-mixed core radius         & fm$_{\rm{radius}}/R_\odot$                 & 0.939 & 0.968& 0.969& 0.961& 0.961& 0.935& 0.938& 0.962& 0.962 \\ 
Fully-mixed core mass           & fm$_{\rm{mass}}/\Mo$                       & 0.885 & 0.942& 0.942& 0.930& 0.930& 0.870& 0.881& 0.930& 0.931 \\ 
 \hline
\end{tabular}
    \caption{Explained Variance scores for tests outlined in Table \ref{tab:testDesc}. We selected 256 tracks (7946 models) quasi-randomly from the grid to form a dev set. The rest of the 227843 main-sequence models were used as the training set.}  
    \label{tab:SPIEV}
\end{table*}

  Table \ref{tab:SPIEV} indicates that by fitting the local frequency ratios (T-2a, T-2b, T-3a, T-3b), we improve the RF's inference on the CBM properties.
  For T-2a and T-2b, the feature importances (not shown here) indicate that the linear intercepts are particularly insightful.
  We note that previously \citet{1990Natur.347..536E} have used the intercepts and gradients of the small separation to identify neutrino physics as the source of the solar neutrino problem rather than exotic stellar physics. Comparisons between fitting as a function of frequency and radial order suggest we  recover a similar amount of information in both instances, with the later having the advantage of helping to mitigate against the surface effect.  Interestingly we note T-4a offers significantly better insight into the helium abundance than the other features. This is likely due to the principal components capturing the variance imparted by the glitch signal. Nevertheless, we find that T-2b offers the best trade off in terms of accuracy for the number of measurements required, whilst also mitigating any bias introduced from improper modelling of the near surface layers.

%As a point of comparison, the mass and radius scaling relations \citep{2009MNRAS.400L..80S} yield an explained variance scores of V$_{\text{e}} =-2.72$ and V$_{\text{e}}=0.69$ respectively for the same training data. 
%The reason for their low scores is because of the wide variety of structures in the \SPI training data that deviate from solar homology. 
%The scaling relations were calibrated in a region of the parameter space more solar-like in physics and structure.
In Figure \ref{fig:stipdists} we provide an alternative measure of the improvement offered by the new diagnostics. 
We plot the error distributions from the dev set for selected parameters. 
The violin plots here indicate the bias and variance (accuracy and precision) in the predictions. 

In general the new input features (T-2a through to T-5b) offer improvement in predictive power over using our standard set of global observables (T-1).
We also find the new CBM parameters are recovered with high fidelity -- particularly the physically meaningful core sizes. 
Whilst it is necessary to see such significant improvement on withheld training data (dev set), true validation is best realized through an independent test set.
For this purpose we benchmarked the new features on the Sun-as-a-Star as well as 16~Cyg~A and B. 
Unfortunately, our new features overfit the training data with less accuracy attained than from using global features (T-1).
The new features are seemingly too sensitive to the perfect model data. 
In order to mitigate this overfitting we have introduced random noise into the  training data when fitting the gradients and intercepts (T-2b). 
This permits accurate inference on stellar parameters whilst still exploiting the insight into overshoot that the new features offer (i.e, we trade off a slightly higher explained variance score for perfect model data in order to make more accurate real-world predictions). %We will seek to optimize the level of noise, and the manner in which it is introduced in future work. 

\iffalse
\begin{figure*}
    \centering
    \includegraphics[width=0.48\textwidth]{FVa.pdf} 
       \includegraphics[width=0.48\textwidth]{FVb.pdf} \\
    \includegraphics[width=0.48\textwidth]{FVc.pdf} 
       \includegraphics[width=0.48\textwidth]{FVd.pdf} \\
           \includegraphics[width=0.48\textwidth]{FVe.pdf} 
       \includegraphics[width=0.48\textwidth]{FVf.pdf} \\
    \caption{Validation set error distributions for the specified stellar parameters. Here we compare the errors in the predictions using the ratios as input features (Test 1, yellow), the gradients and intercepts of the ratio profiles (Test 2a, green) and taking the principal components of the frequencies (Test 4a, blue).}
    \label{fig:stipdists}
\end{figure*}
\fi

\begin{figure*}
    \centering
    \includegraphics[width=0.48\textwidth]{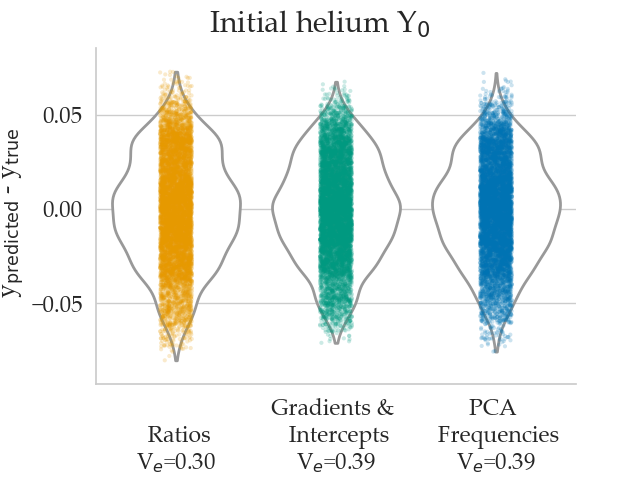} 
       \includegraphics[width=0.48\textwidth]{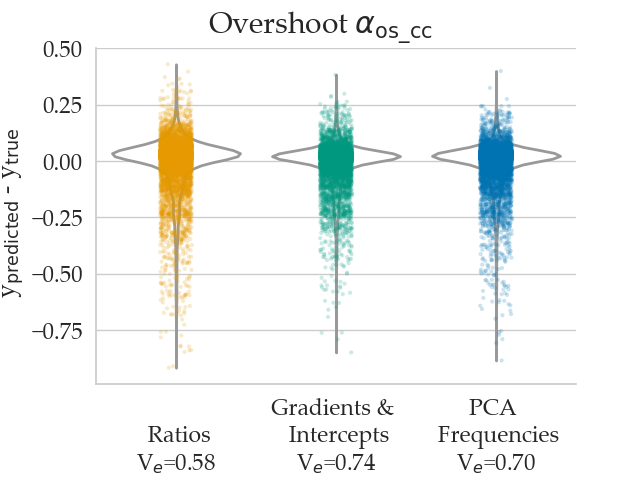} \\
    \includegraphics[width=0.48\textwidth]{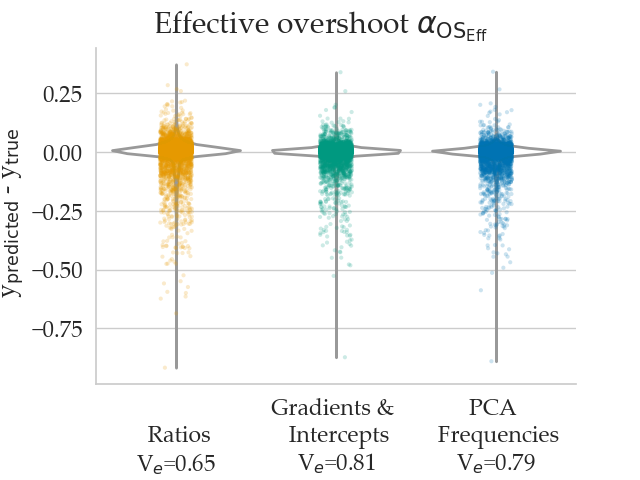} 
       \includegraphics[width=0.48\textwidth]{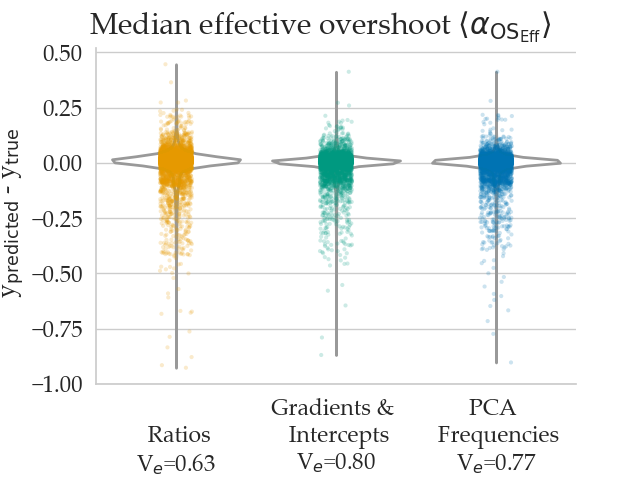} \\
           \includegraphics[width=0.48\textwidth]{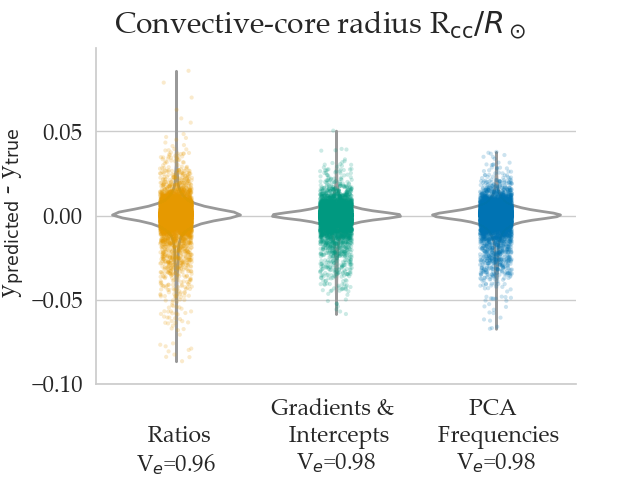} 
       \includegraphics[width=0.48\textwidth]{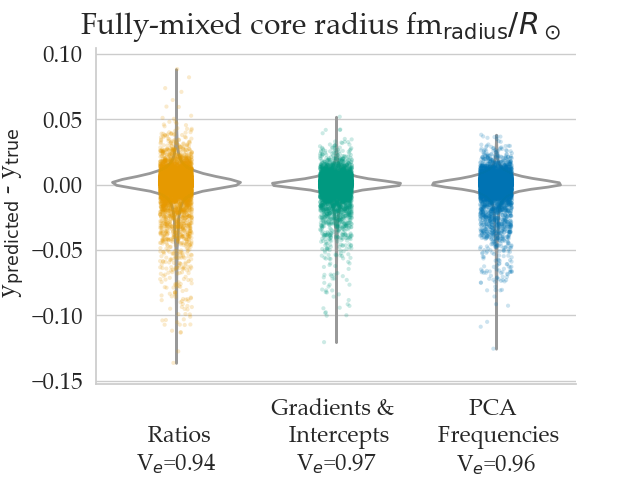} \\
    \caption{Development set error distributions for the specified stellar parameters. Here we compare the errors in the predictions using the ratios as input features (T-1, yellow), the gradients and intercepts of the ratio profiles (T-2a, green) and taking the principal components of the frequencies (T-4a, blue).}
    \label{fig:stipdists}
\end{figure*}
\

\subsubsection{Pre-processing Classifiers}
When using machine learning algorithms it is important to understand their strengths and weakness.
RF's perform particularly well on continuous data but are not well-suited to say fitting seasonal trends in timeseries data \citep{hastie2005elements}.
Many of the new CBM parameters do not form continuous distributions; they are zero in the absence of a convective core.
The RF, however, will regress across models with convective and stable cores attributing non-zero, albeit small, values for the CBM quantities for stars with radiative cores.

In order to mitigate this effect we have trained a classifier to determine whether the target has a convective core. 
We evaluated several classification algorithms (e.g., decision trees, Gaussian processes, support vector machines, neural networks, to name but a few) and found random forests best suited in terms of speed and accuracy.     
We attain an accuracy of 97\% on withheld training data when using the global features (T-1). 
The classifier evaluates whether a set of observables corresponds to a convective core at the current age.
If a given realization is deemed radiative we set the relevant CBM parameters to zero. 
If it is classified as having a convective core, we regress for the current-age CBM parameters. 
%An example can be seen in Figure XXX where we have calculated the convective core radius and and effective overshoot from 10,000 realizations of Sun-as-a-Star data -- a star we know for sure does not have a convective-core. 
These steps ensure that regression is applied appropriately throughout the parameter space and allows for a better representation of models with extreme physics that induce convection in the core. 
As the classifier is pre-trained, it adds negligible computational time for the pipeline. 

Ultimately we would like to rapidly and robustly analyse stars in all phases and stellar systems with \textsc{SPI}. 
Whilst training an all-encompassing grid of models is one possible strategy, it is important to remember that not all input features provide insight in all evolutionary phases (i.e., the small separation loses its diagnostic potential in subgiants \citealt{2011ApJ...742L...3W}). Furthermore, not all quantities can be defined in every system. 
We can improve the accuracy of \SPI if \textit{a priori} knowledge of the evolutionary stage can be supplied. 
This ensures that the input features and training data are exploited optimally. 

We have implemented a RF classifier in the \SPI pipeline to determine the evolutionary phase based on  $\nmx$, $\Teff$, $\Dnu$, and $\Fe$. Evolutionary phases were divided according to three sampling regimes listed in \S \ref{sec:SPIov}. 
Detailed results are presented in Appendix \ref{sec:EPC}, however, we note here that the classifier accurately distinguishes between main sequence and subgiant stars thus ensuring with high probability that the correct phase is included in the training data.

\section{Observational Surveys and Targets}
\label{sec:Targets}
\begin{figure*}
    \centering
    \includegraphics[width=\textwidth]{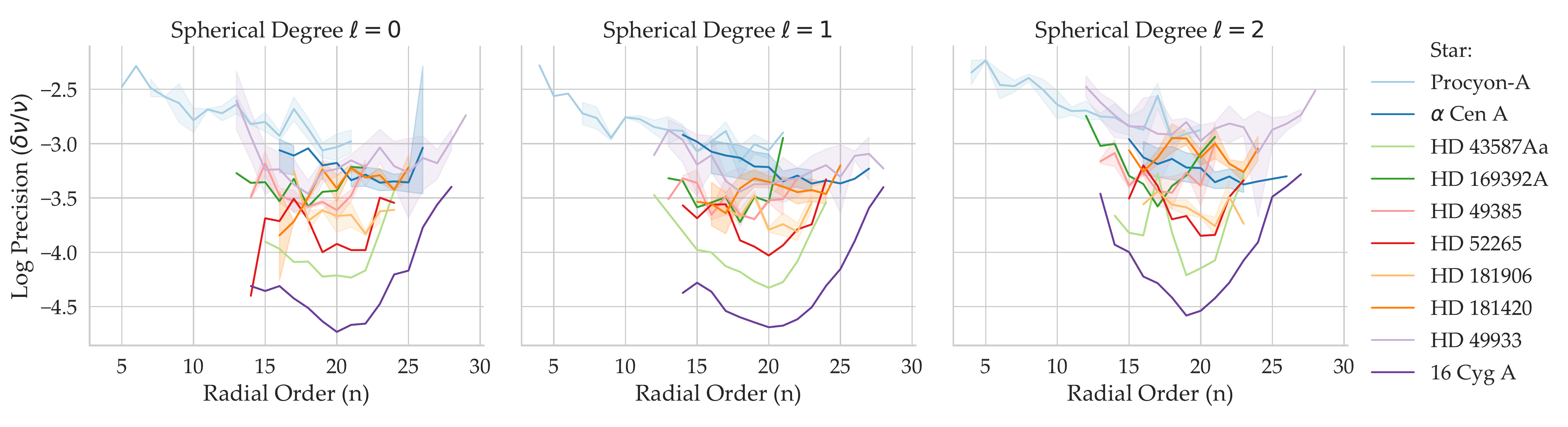}
    \caption{The frequency precision for oscillation modes with different spherical degrees as a function of radial order. Here we compare seven \corot \ stars, $\alpha$~Cen~A and Procyon~A with the \Kepler \ Star 16~Cyg~A. Shaded regions indicate the uncertainty stemming from different mode identifications in the oscillation spectra (see Table \ref{tab:corSum}).} 
    \label{fig:precisions}
\end{figure*}

\begin{table*}
    \centering
        \caption{An overview of  \corot \ and radial velocity stars analyzed in this study.}

    \begin{tabular}{lcccccC{2cm}c}
    
\multicolumn{1}{>{\centering}p{1cm}}{~\\Star ID} & \multicolumn{1}{>{\centering}p{1cm}}{Spectral \\ Class} & \multicolumn{1}{>{\centering}p{1cm}}{~\\$\Teff$(K)} &\multicolumn{1}{>{\centering}p{1cm}}{~\\$\Fe$} &  \multicolumn{1}{>{\centering}p{0.75cm}}{ $\Dnu$ \\ ($\mu$Hz)} &  \multicolumn{1}{>{\centering}p{4.5cm}}{ Mode identification \\ (as per reference) }  & \multicolumn{1}{>{\centering}p{2cm}}{ Time Series Duration} &\multicolumn{1}{>{\centering}p{2cm}}{ In-Text \\ Designation$^*$}   \\ \hline \hline 
\multicolumn{8} {c} {\bfseries \corot Oscillation Spectra}          \\                              \hline
HD 43587Aa                & G0        & 5900          & -0.02     & 107               &{\citet{2014A&A...564A..34B}}  & 145 days               & HD43587Aa-$\nu 1$  \\[3pt] 
HD 169392A$^{\dagger}$  & G0        & 5985          & -0.07     & 56                &{\citet{2013A&A...549A..12M}}  & 91 days                & HD169392A-$\nu 1$  \\[3pt]
\multirow{ 2}{*}{HD 49385$^{\dagger}$} & \multirow{ 2}{*}{G0} & \multirow{ 2}{*}{6095} & \multirow{ 2}{*}{0.09} & \multirow{ 2}{*}{56}       &{\citet{2010A&A...515A..87D}} Scenario 1   & 137 days & HD49385-$\nu 1$\\
                                       &                      &                        &                        &                            & {\citet{2010A&A...515A..87D}} Scenario 2  & 137 days & HD49385-$\nu 2$  \\[4pt]
HD 52265                & G0        & 6100          & 0.2       & 98                &{\citet{2011A&A...530A..97B}}  & 117 days               & HD52265-$\nu 1$ \\[3pt]
\multirow{ 2}{*}{HD 181906}  &\multirow{ 2}{*}{F8}   & \multirow{ 2}{*}{6300} & \multirow{ 2}{*}{-0.11}   & \multirow{ 2}{*}{88}             & {\citet{2009A&A...506...41G}} Scenario A  & 156 days  & HD181906-$\nu 1$ \\
                                       &                      &                        &                        &                            & {\citet{2009A&A...506...41G}} Scenario B   & 156 days & HD181906-$\nu 2$ \\[4pt] 
\multirow{ 2}{*}{HD 181420} & \multirow{ 2}{*}{F2}            & \multirow{ 2}{*}{6580} & \multirow{ 2}{*}{0.0}     & \multirow{ 2}{*}{75}    &{\citet{2009A&A...506...51B}} Scenario 1 &  156 days   & HD181420-$\nu 1$ \\
                                       &                      &                        &                        &                            & {\citet{2009A&A...506...51B}} Scenario 2&  156 days   & HD181420-$\nu 2$ \\[4pt] 
\multirow{ 5}{*}{HD 49933}  & \multirow{ 5}{*}{F5}            & \multirow{ 5}{*}{6770} & \multirow{ 5}{*}{-0.37} & \multirow{ 5}{*}{85}      & {\citet{2009A&A...506...15B}} M$^2_A$  & 60 days  & HD49933-$\nu 1$ \\
                                       &                      &                        &                        &                            &{\citet{2009A&A...506...15B}} M$^2_B$   & 60 days & HD49933-$\nu 2$ \\
                                       &                      &                        &                        &                            &{\citet{2009A&A...507L..13B}}& 180 days  & HD49933-$\nu 3$ \\
                                       &                      &                        &                        &                            &{\citet{2008A&A...488..705A}}  & 60 days  & HD49933-$\nu 4$ \\
                                       &                      &                        &                        &                            &{\citet{2009A&A...506.1043G}} & 60 days  & HD49933-$\nu 5$ \\[4pt] \hline \hline
\multicolumn{8} {c} {\bfseries \corot Asteroseismic Global Parameters}          \\            \hline                  
HD 46375                & K0        & 5300          & 0.39      &153 & {\citet{2010A&A...524A..47G}} & 34 days & \\[4pt]
HD 175726               & F9/G0     & 6000          & -0.22     & 97 & {\citet{2009A&A...506...33M}} & 27 days & \\[4pt]
HD 170987               & F5        &6540           & -0.15     &55 & {\citet{2010A&A...518A..53M}} &149 days &\\[4pt]
HD 175272               & F2        &6675           & 0.08      & 75 &{\citet{2013A&A...558A..79O}} & 27 days &\\[4pt] \hline \hline
\multicolumn{8} {c} {\bfseries Radial Velocity Oscillation Spectra}       \\      \hline                        
\multirow{ 2}{*}{Procyon~A} & \multirow{ 2}{*}{F5}        &\multirow{ 2}{*}{6575}     &  \multirow{ 2}{*}{0.0} & \multirow{ 2}{*}{54} & {\citet{2010ApJ...713..935B}} Scenario A &  Multi Site RV & ProcyonA-$\nu 1$ \\
                                       &                      &                        &                        &                            &{\citet{2010ApJ...713..935B}} Scenario B $^{\ddagger}$ &  Multi Site RV & ProcyonA-$\nu 2$ \\[4pt]
\multirow{ 3}{*}{$\alpha$~Cen~A} & \multirow{ 3}{*}{G2}        &\multirow{ 3}{*}{5800}     &  \multirow{ 3}{*}{0.23} & \multirow{ 3}{*}{106} & {\citet{2007A&A...470..295B}} & HARPS RV & $\alpha$CenA-$\nu1$ \\

                                       &                      &                        &                        &                            &{\citet{2004ApJ...614..380B}} & UVES, UCLES RV&  $\alpha$CenA-$\nu 2$ \\ 
                                       &                      &                        &                        &                            &{\citet{2010A&A...523A..54D}} &CORALIE, UVES, UCLES RV&  $\alpha$CenA-$\nu 3$ \\                                                                               
                                       \hline
    \end{tabular}

\flushleft
$^*$ In the supplementary online material. \\
$^{\dagger}$ denotes that the star is likely a subgiant. \\
$^{\ddagger}$ We have used these frequencies but relabelled the radial order based on an expected regular pattern. 
    \label{tab:corSum}
\end{table*} 

The new SPI input features have been designed to better extract information from the stellar interior. 
Moreover, in this work one of the aims is to determine how asteroseismic precision impacts the determination of key CBM properties, which amongst other things is crucial for reliable stellar ages.
To this end we have reprocessed the results from \citet{2019A&A...622A.130B} with the new pipeline which include data from 
the KAGES \citep{2015MNRAS.452.2127S,2016MNRAS.456.2183D} and Legacy \citep{2017ApJ...835..172L,2017ApJ...835..173S} surveys. 

A majority of stars in the \TESS field will be observed for 27 days; far shorter than the four-year timeseries of \Kepler.
The duration and cadence will reveal ridges in the echelle diagram although not with the precision required to probe the interior in detail.  
Many targets, however, will lie in the repeated/continuous viewing zones resulting in timeseries longer than 54 days.
For these stars we can expect a comparable frequency resolution to previously studied radial velocity targets (e.g., Procyon~A, $\alpha$~Cen~A) or solar-like oscillators in the \corot \ field which we analyse with \textsc{SPI} below.
We summarize the main properties of these stars in  Table \ref{tab:corSum}, while in Figure \ref{fig:precisions} we compare their measurement precisions with the precision obtained for 16~Cyg~A -- one of the \Kepler \ benchmarks.
The combination of different surveys, precisions and observations essential for our scientific goal.

\section{Results}
\label{sec:results}

\subsection{Summary of Star-to-Star validation}

As a final form of validation for the updated pipeline, we compare the \corot and RV results from \SPI with  \textsc{Unified tool for Distance, Age and Mass estimation} \citep[\textsc{UniDAM},][]{Mints2017, Mints2018}\footnote{\url{http://www2.mps.mpg.de/homes/mints/unidam.html}}, and previous studies in the literature.
The \unidam solutions  provide a complementary homogeneous analysis\footnote{We also refer the reader to  \citet{2019MNRAS.489.1753Y} which was published as we were concluding this work.} based on parallaxes and isochrone fitting. 
Key findings are summarized below with detailed comparisons available in the supplementary material. 

Despite the different methodologies, we find \SPI and \unidam yield consistent masses, radii and luminosities for the \corot \ and RV targets.
This is encouraging given the large asteroseismic uncertainty associated with these stars.
In most cases \SPI identified the solution found by previous studies upon adopting their assumptions and observational constraints.
As we processed most of the available frequency lists we also identified several lesser minima for some targets. 
Although stellar ages are code dependent, they generally agree within the (large) uncertainties. 
We highlight the following noteworthy results from the comparison tests:

\begin{itemize}
    \item \textbf{HD 43587Aa:} The asteroseismic radius (\SPI, \citealt{2014A&A...564A..34B}) is only marginally consistent with the interferometerically determined value. Sensitivity of limb darkening models is an obvious starting point for further investigations. 
    \item \textbf{HD 49385:}  \citet{2011A&A...535A..91D} identified high and low overshoot solutions for this star. \SPI strongly favours a low overshoot efficiency. 
    \item \textbf{HD 52265:} Both \SPI and \citet{2014A&A...568A.123B} (who used the same evolution code but applied a surface correction to match individual frequencies)  infer a highly sub-solar mixing length for this archetypal G-type star.
    \item \textbf{HD 181906:} This star is far from the ideal benchmark target. It is an F-star with low power excess, it is comparatively low-metallicity, exhibits strong magnetic activity, rotates 10 times faster than the Sun, and its lightcurve is contaminated by a binary companion. Adoption of Gaia lightcurves (uncorrected for binaries) significantly skews the inferred properties of this star whilst the use of a corrected luminosity \citep{2009A&A...506..235B} differs by over 2$\sigma$ compared to the asteroseismically inferred value.
    \item \textbf{Procyon~A:} \citet{2014ApJ...787..164G} used Bayesian evidence to determine overshooting by $1.0-1.5H_P$ results in the best fitting model for this star. The efficiencies inferred by \SPI are up to a factor of 50 less when only radial modes are considered. There are large variations in the predictions from different studies/methods for this star.  
    \item $\mathbf{\alpha}$\textbf{~Cen~A:} The convective core classifier determined 84\% of realizations for this star posses a convective core. This is similar to the probability densities from  \citet[][77\%]{2018MNRAS.479L..55N} and \citet[][89\%]{2016MNRAS.460.1254B} (with overshoot and diffusion included in their models). 
\end{itemize}

\subsection{CBM properties from pressure-mode asteroseismology}
\begin{figure*}
    \centering
    \includegraphics[ width=\textwidth]{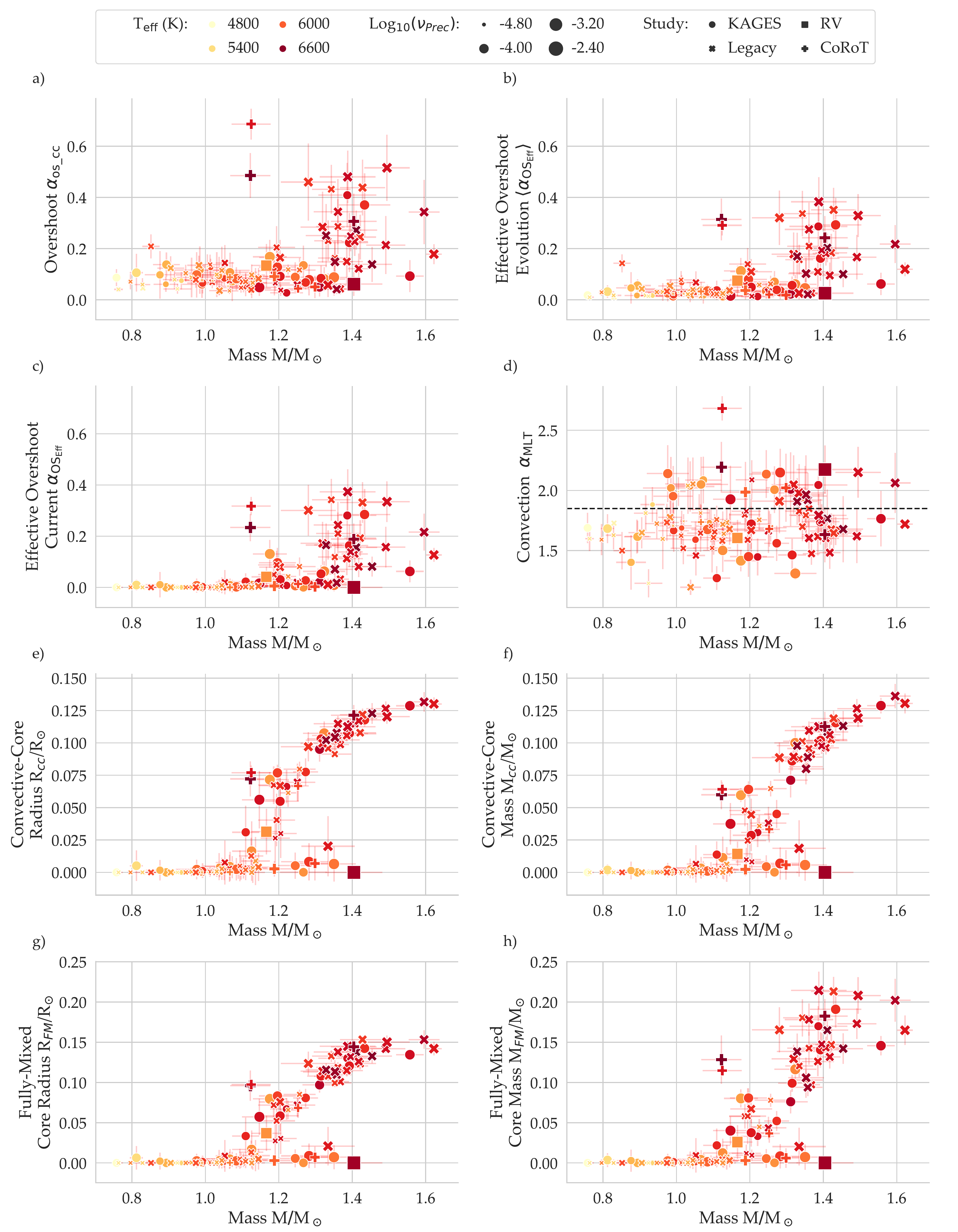}
    \caption{CBM parameters as function of stellar mass for asteroseismic targets. The indicative precision on the frequencies is determined by taking an average of the three closest modes around $\nmx$. The dashed line in the $\alpha_{\mathrm{MLT}}$ plot indicates the solar calibrated value.}
    \label{fig:OScomparison}
\end{figure*}

In Figure \ref{fig:OScomparison} we show SPI predictions for selected CBM parameters as a function of mass for \Kepler, \corot \ and RV stars. 
Explained variance scores indicate that overshoot efficiencies can be inferred with accuracy sufficient enough  to identify outliers and large-scale trends (Table \ref{tab:SPIEV}, T-2b). 
Like many of the current-age stellar parameters (mass, radius, luminosity), the convective- and fully-mixed cores can be predicted with high fidelity and we consider 
SPI inferences accurate.

In Panels \ref{fig:OScomparison}a-\ref{fig:OScomparison}c we plot different measures of the overshoot efficiency. 
Panel \ref{fig:OScomparison}a indicates overshoot  ($\alpha_{\mathrm{os\_cc}}$) as set in the \textsc{MESA} parameter file. 
This is compared to the pseudo `effective overshoot' in Panel \ref{fig:OScomparison}b, which we have determined through our modifications to the evolution code. 
Recall that this is a median of the effective overshoot efficiencies along an evolutionary sequence.
In Panel \ref{fig:OScomparison}c we plot the inferred current-age overshoot -- the instantaneous effective overshoot for a given model. 
As indicated by Figure \ref{fig:OSIO}, and by comparing Panels \ref{fig:OScomparison}a with \ref{fig:OScomparison}b or \ref{fig:OScomparison}c, it is clear that the efficiency that the code effectively overshoots is often not nearly as extreme as that set by the user.

For most of the stars in the sample, the inferred CBM parameters are consistent with the range of values determined from calibrations to binaries and clusters. The inferred overshoot efficiencies do not display a clear dependence on stellar mass.
However, our overshoot predictions are characterised by large uncertainties.  They reflect the systematics in the modelling, propagated through widely varying the physics in the training data as well as the degeneracies in the parameter space. 
We note that two \corot \ targets, namely HD~181906 and HD~49933, appear as outliers in (all) the panels.  
These stars are the focus of \S \ref{sec:CL}, although we note here that their large asteroseismic uncertainty is a necessary but insufficient condition for their anomalous CBM parameters. 
We find in general that the mean and the scatter in the effective overshoot increase substantially beyond $M \gtrapprox 1.3 \Mo$. These stars are characterised by thin convective envelopes and larger asteroseismic uncertainty -- a point we return to in \S \ref{sec:IGR}

In the remaining panels the reported parameters refer to their current age values. 
In Panels  \ref{fig:OScomparison}e-\ref{fig:OScomparison}h we separate out the mass and radii of the convectively unstable core as well as the the fully-mixed region. 
Ultimately the size of the convective core depends on the stellar mass, metallicity and central-hydrogen abundance. 
Save for the two outliers, the convective-core sizes naively follow the expectations of stellar evolution and their spectral type:
The hotter, more-massive main-sequence stars possess larger convective cores whilst inert centres are inferred for the subgiants. 
%Interpreting the size of the fully-mixed region introduces an additional complication from the effective overshoot efficiency.
For stars which are not near to core-hydrogen depletion ($X_c > 0.3$), we find that the mass of the fully mixed core is reasonably well-modeled ($R^2 = 0.95$) as a sigmoidal function of the stellar mass (see Figure~\ref{fig:sigmoid}).

In Panel  \ref{fig:OScomparison}d we plot the convective efficiency ($\alpha_{\mathrm{MLT}}$) inferred by SPI, with the solar-calibrated value indicated by the horizontal line. 
The SPI results demonstrate a flat distribution with no clear trend with temperature/spectral type (cf. \citealt{2015A&A...573A..89M, 2014MNRAS.442..805T}).
This should be interpreted with the caveat that the RF has a qualitative understanding of the mixing length parameter but is unable to predict with the same high-fidelity as the current-age stellar properties \citep{2017ApJ...839..116A}. 
Nevertheless, we indeed find qualitatively similar results to previous studies\footnote{See supplementary material for our star-by star-comparison. In particular the case of HD 52265 where \citet{2014A&A...568A.123B} used the same evolution code but adopted the strategy of optimization and frequency matching.} in that most stars are best characterized by a sub-solar mixing length. The mixing length adds an extra degree of freedom in model fitting procedures and can act as a fine tuning parameter. Recent work by \citet{2019MNRAS.484.5551J} and \citet{JA19} have successfully coupled the stratifications from 3D atmospheres to 1D structures removing the need to include $\alpha_{\mathrm{MLT}}$ as a free dimension (see also the method by \citealt{2018ApJ...869..135S}). In coupled models $\alpha_{\mathrm{MLT}}$ only needs to bridge the entropy difference of a very shallow layer between the 3D envelopes and 1D structure rendering the models essentially insensitive to this parameter.

 \subsection{A closer look at HD~49933 and HD~181906}
\label{sec:CL}
\iffalse
Figures \ref{fig:OSIO} and \ref{fig:OScomparison} demonstrate that there can be significant differences between the specified overshoot and the actual extent of mixing in the code. 
The manner in which the code truncates the overshooting distance goes a long way to explaining why some models require seemingly extreme values in order to match observations.
We find upon calculating an `effective overshoot' throughout the evolution, the required efficiencies are comparable with values determined from calibrations to clusters and binary systems --  a naive application of the overshoot formalism would result in core sizes $>30\%$ larger than those currently inferred. 
However our analysis has identified two stars that remain clear outliers in Figure \ref{fig:OScomparison}, namely the \corot \ targets HD~49933 and HD~181906.
We note that of the stars with a convective core, they are measured with the lowest precision (Figure \ref{fig:precisions}).  
\fi

We have already outlined some of the reasons why HD~181906 is such a difficult target to study. 
Undaunted by the inherent challenges, nine groups analysed this star fitting modes across seven radial orders in a test of their peak-bagging pipelines \citep{2009A&A...506...41G}. 
They arrived at two possible mode identifications which we processed with SPI. 
Both frequency tables favour input overshoot values in the range of $\alpha_{\mathrm{os\_cc}}=0.65-0.70H_P$ or a median effective overshoot  $\left<\alpha_{\rm{OS}_{\rm{Eff}}}\right>\approx 0.3H_P$. 
We note that due to the limited number of modes identified, we could not utilise the gradients and intercepts (T-2b parameters) to diagnose the CBM properties. 
Rather we had to rely on the global parameters outlined in Table \ref{tab:testDesc} (T-1).

HD~49933 is  another challenging star in the \corot \ field having proven itself contentious for observers and modellers alike.
On account of the large line widths, mode identification is difficult. Initially, it was  unclear as to whether certain peaks in the power spectrum correspond to closely spaced pairs of $\ell =0$ and $\ell =2$ modes or broad $\ell=1$ modes.
The uncertainties motivated a series of papers \citep{2008A&A...488..705A, 2009A&A...507L..13B, 2009A&A...506...15B, 2009A&A...506....1A, 2009A&A...506.1043G} which have sought to disentangle the issue of mode identification through various mathematical and statistical techniques.  The provided frequency lists were subsequently used by several groups in order to characterize HD~49933 in detail. \citet{2014ApJ...780..152L} and \citet{2015A&A...574A..45R} generated large grids of models with different input physics and demonstrated that a wide spread of models
are consistent with the data. The latter explicitly concluding the precision on the frequencies  (Figure \ref{fig:precisions}) is not good enough to constrain the large number of possible models.

Here we wish to highlight results by \citet{2014ApJ...780..152L} who used YREC models to characterise HD~49933.
Their best-fit model required an input overshoot of $\alpha_{\mathrm{os\_cc}}=0.6$. 
This value is in fact similar to the efficiency determined by \textsc{SPI}, which in the case of \textsc{MESA} corresponds to an effective overshoot of $\left<\alpha_{\rm{OS}_{\rm{Eff}}}\right> \approx 0.3 H_P$. 
\textsc{YREC}, like \textsc{MESA} and \textsc{GARSTEC} truncates overshoot based on the relative radial extent of the convective core and proposed overshoot region. 
The YREC truncation algorithm may also somewhat help explain the extreme overshoot ($\alpha_{\mathrm{os\_cc}}=1-1.5H_P$) needed  by \citet{2014ApJ...787..164G} to fit Procyon~A (as may some of the reasons below).

\begin{figure*}
    \centering
    \includegraphics[width=\textwidth]{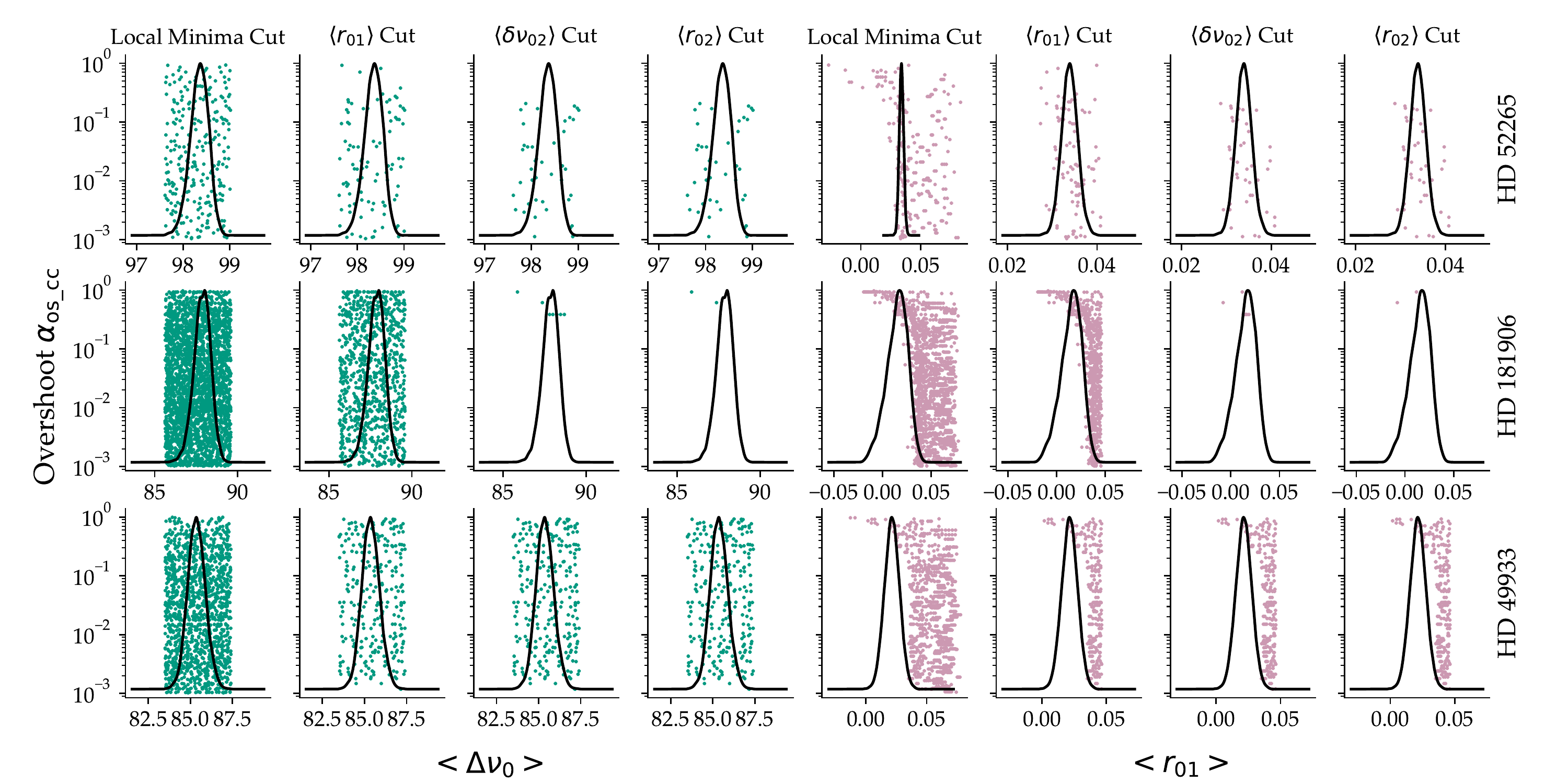}
    \caption{Models in the $\Dnu - \alpha_{\mathrm{os\_cc}}$ plane (left, green) and the  $\left<r_{01}\right> - \alpha_{\mathrm{os\_cc}}$  plane (right, pink) after sequential cuts to the training grid.  In addition to the remaining models, we plot the ranges $\Dnu$ and $\left<r_{01}\right>$ can take following the perturbation procedure (respective black distributions).}
    \label{fig:parm}
\end{figure*}

\begin{figure}
    \centering
    \includegraphics[width=\columnwidth]{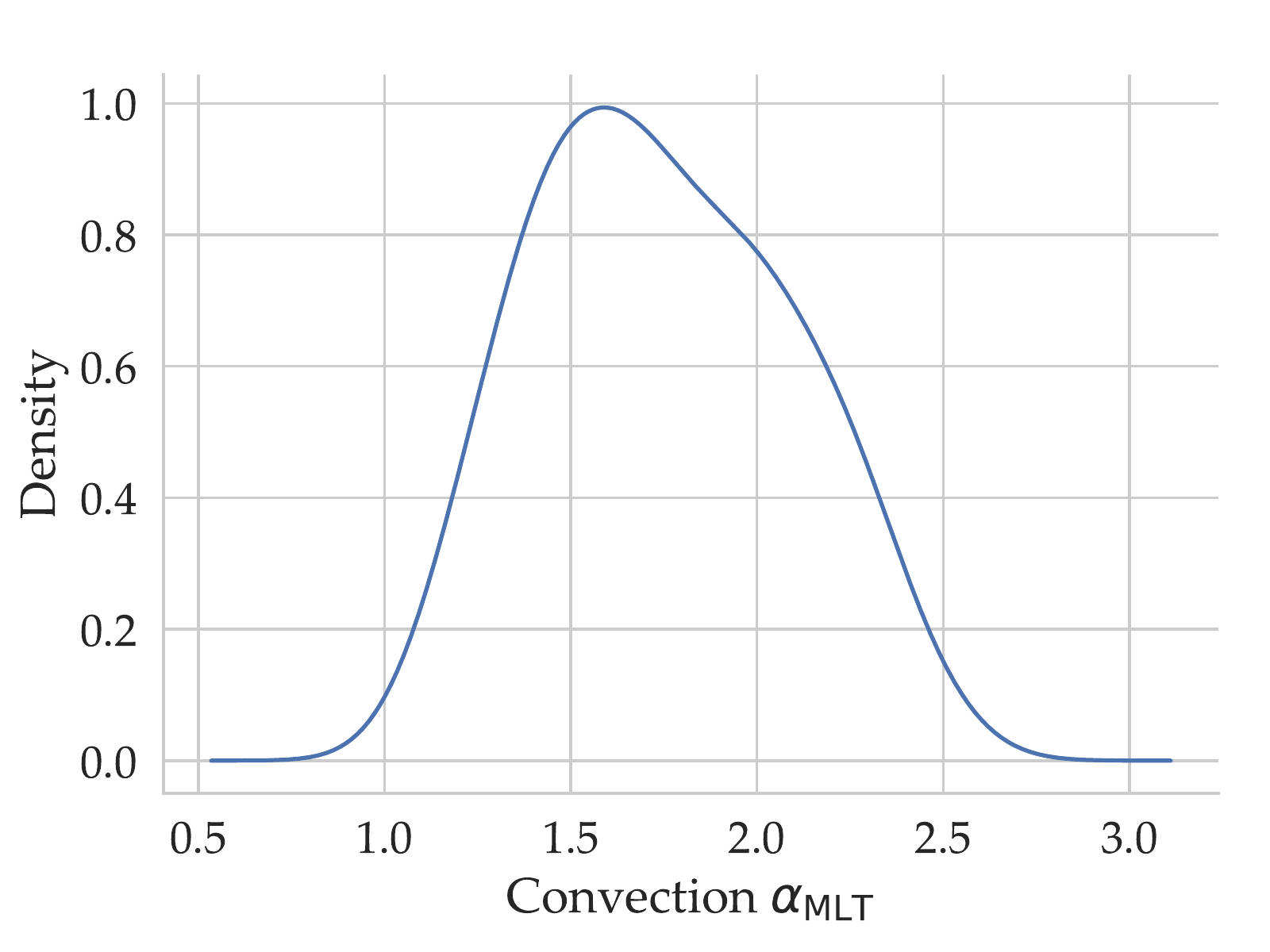}
    \caption{Distribution of $\alpha_{\mathrm{MLT}}$ for the 55 remaining models consistent with HD~52265 following parameter cuts in Figure \ref{fig:parm}.}
    \label{fig:HDMLT}
\end{figure}

The amount of overshoot required to match HD~49933 and HD~181906 is higher than other stars of similar mass.
Their resultant convective- and fully-mixed core sizes are also larger than average.
Understanding what is driving the asteroseismic solution (e.g. \SPI and \citealt{2014ApJ...780..152L}) is prudent as we can expect data for similar targets and of similar quality in the \TESS survey. 
The need for such large core masses naively points to one of two scenarios. 
Either there models are missing essential physics which the high overshoot is trying to compensate for, or there is an issue with the mode identification which is demanding extreme structures.  

We first inspected the RF's decision pathways for these stars. 
The final overshoot value are very much reliant on splits in the $\left<r_{01}\right>$, $\left<r_{10}\right>$ and $\left<\Dnu\right>$ parameter space.
The decision trees hinted that the RF is having difficulty finding solutions for HD~49933 and HD~181906 that simultaneously explain  $\left<r_{01}\right>$ and $\left<\Dnu\right>$  with low overshoot. 

Through Figure \ref{fig:parm} we demonstrate how a grid-based search would encounter the RF's quandary. 
Essentially  $\chi^2$  searches are governed by seismic constraints as they are measured with higher precision than spectroscopic observables. 
The optimization algorithms normally do not indicate the elements that were important for arriving at a given minima. 
Here we perform `optimization by hand' to identify models in a grid that an algorithm would select as a best fit to these stars.

We use the RF's training data to perform a pseudo grid search in order to understand why the asteroseismic solutions tend towards high overshoot.
We search for suitable models that reproduce the observable properties of HD~49933, HD~181906 and HD~52265. 
The latter a \corot \ star also measured with low precision but with inferred CBM parameters that are considered typical. 
We determine suitable ranges for each observable by running the stars through the  \SPI perturbation procedure. 

In Figure \ref{fig:parm} we follow the hints left by the RF decision pathways. 
In the four left-most columns (green points) we plot the $\Dnu - \alpha_{\mathrm{os\_cc}}$ plane of the model parameter space, whilst in the four right-most columns we consider the $\langle r_{01}\rangle - \alpha_{\mathrm{os\_cc}}$ plane (pink points). In each successive column we make cuts to the grid, filtering out models that are inconsistent with a distinct set of observable properties.  
We first make a cut to the parameter space in $\Teff$, $\Fe$ and $\Dnu$ as one might do when identifying a local minima region in a grid search.
Subsequent cuts are then made to models inconsistent with measured  $\left<r_{01}\right>$, $\left<\delta \nu_{02}\right>$ and $\left<r_{02}\right>$ values.
In each panel we overplot the distribution for the parameter on the abscissa. This demonstrates the degree to which the models remaining in the grid are consistent with that observable quantity.
Note that the order of the parameter cuts (frequency separation followed by frequency ratio) helps to diagnose what role the surface effect plays in driving the asteroseismic solution. 

The three stars each tell a different story. 
For HD~52265 several models exist, covering a range of overshoot efficiencies, that are consistent with all the observational constraints.

We find that for HD~181906, the local minima cut is not very restrictive. 
With the subsequent $\left<r_{01}\right>$ cut we still find many models simultaneously consistent with $\left<\Dnu\right>$ and $\left<r_{01}\right>$ but we note this is by virtue of most models falling within the upper tail of the $\left<r_{01}\right>$  distribution. Once we introduce constraints from the small separation, only a handful of models -- all with high overshoot, remain consistent with $\left<\Dnu\right>$ and $\left<r_{01}\right>$. This is irrespective whether we apply a $\left<r_{01}\right>$, $\left<r_{02}\right>$, or  $\left<\delta \nu_{02}\right>$ cut first. 
This suggests it is not a case of parameters with different sensitivities to the near surface layers placing incompatible constraints on the structure. 
Here we can rule out the surface effect as the primary driver of the asteroseismic solution, rather the central frequencies of the $\ell=0$ are incongruent with the $\ell=1$ and $\ell=2$  modes. Here the $\ell=2$ modes are seemingly the most troublesome.

For HD~49933 there are many models consistent with the large separation once the parameter cuts have been made.
Here the uncertainty on the $\ell=2$ modes renders their constraints on the structure not overly restrictive. 
For this star it is the $\ell=1$ modes that are telling. 
We find many acceptable models, varied in overshoot efficiency however they correspond to the high end tail of the $\langle r_{01}\rangle$ measurement.
Whilst at the centre of the  $\langle r_{01}\rangle$  distribution only high overshoot models are possible. 
It is the models located near the centre of the  $\langle r_{01}\rangle$  distribution that will return the lowest  $\chi^2$ for a grid search.
In terms of the RF, the distribution reflects the relative number of realizations determined from the frequency measurements and their uncertainties. 
Thus the RF will also use those models to infer high overshoot when constructing its decision trees. 
%However we can not rule out the possibility that a surface correction may shift lower overshoot models accordingly. 
Interestingly,  \cite{2015A&A...574A..45R} found many solutions with $\chi^2 < 1$ in his analysis of HD 49933.
His method relied on finding potential structures based on luminosity and his phase-matching procedure, the latter which  mitigates against the impact of the improper modelling of the near-surface layers.
His search yielded dozens of probable models that varied in their overshoot efficiency but he did not use $\Dnu$ as a constraint.
Similarly, it can be seen from Figure \ref{fig:parm} that omitting $\left<\Dnu\right>$ and using $\left<r_{01}\right>$ as a constraint (which also mitigates against the surface term), allows us to identify many probable models varied in their overshoot efficiency.  
%It is clear that for HD~181906 the radial modes were incongruent with the non-radial modes. 
For HD~49933 there is an inconsistency between the $\ell=0$ and $\ell=1$ modes driving the asteroseismic solution to high overshoot (as the RF indicated), however, our current analysis is unable to quantify the role of surface term in causing this. 
%We also find the uncertainty on the $\ell=2$ modes too large to 

\section{Discussion}
\label{sec:discussion}
\subsection{Asteroseismic Inference}
For most stars in this study, we find CBM parameters consistent with expectations from calibration studies.
However, the \corot \ targets  HD~49933 and HD~181906 stand out from their counterparts. 
The asteroseismic solution for these stars tends to higher CBM parameters in order to find structures simultaneously 
consistent with constraints from radial and non-radial modes. 
Whilst there could be stochastic differences in their physical processes or rare events in the evolution that may necessitate larger than typical convective cores, Occam's razor suggests otherwise.
It is perhaps no coincidence that targets in question are both F-stars and suffer from short mode lifetimes and potentially attenuation from magnetic fields \citep[][recall HD~181906 has strong magnetic fields]{2015AdSpR..56.2706B}. The low asteroseismic precision and resultant large linewidths make mode identification difficult, as evidenced by the series of papers on HD 49933 (see  also \citealt{2012ApJ...751L..36W}) and these difficulties have motivated novel methods of analysis such as using mode centroids  \citep{2010ApJ...713..935B, 2019MNRAS.485..560C}.

With this in mind it is prudent that we briefly mention Procyon~A; another F star measured with low precision.
The RF classifier unequivocally infers a subgiant evolutionary state for Procyon~A. 
As a consequence, our pipeline only makes use of radial modes in its analysis (we are yet to implement algorithms that automatically detect and handle mixed modes). 
With the available constraints, we infer an input overshoot of $\alpha_{\mathrm{os\_cc}}=0.05H_P$ or  $\left<\alpha_{\rm{OS}_{\rm{Eff}}}\right> = 0.02 H_P$.
This is compared to best fitting models from  \citet{2014ApJ...787..164G} who require overshooting efficiencies of between $1.0-1.5H_P$. 
Given that we have demonstrated SPI's ability to reproduce the results of other studies, identifying any differences in the respective analyses will no doubt prove telling.
\citet{2014ApJ...787..164G} found that a radiative overshoot region returned a better fit to the data than assuming it is adiabatically stratified. We too assume a radiative overshoot region. 
Crucially, they included fits to the large frequency separation for each of the $\ell=0,1,2$ modes in their likelihood whereas \textsc{SPI} is limited to $\ell=0$.
Thus it is tempting to wonder whether inconsistencies between radial and non-radial modes is yet again demanding extreme structures. 
We note that Procyon~A is featured in the \emph{TESS} input catalog with new measurements to be published in the near future.

Understanding these issues will help with the characterization of \emph{TESS} targets, as we can expect observations of stars of similar spectral type with comparable asteroseismic precision. The correct determination of CBM efficiencies is particularly vital for accurate stellar ages. By utilizing our improvements to SPI, we have demonstrated that the efficiencies of CBM processes in seismic targets are mostly reasonable and consistent with results from calibration studies.

On a final note we wish to return to results for HD~55625 and the many other stars that require sub-solar mixing length to explain their seismic structures. Such results are not unique to \SPI and are commonplace in the literature. One of the advantages of our data-driven method is that we have been able to quantify the accuracy with which we can infer this parameter - a parameter that suffers from several degeneracies with other quantities in stellar evolution. Following our parameter cuts in Figure \ref{fig:parm}, 55 models remained consistent with the observations of HD~55625.  We plot a normalized distribution of their $\alpha_{\mathrm{MLT}}$ parameter in Figure \ref{fig:HDMLT}. The solutions reside in a highly redundant section of the stellar evolution space with various mixing lengths consistent with the observable data. Moreover, we in fact identify an abundance of models with sub-solar mixing-length efficiency. This is compared to  HD~49933 and HD~181906 where only a handful of models with high overshoot are consistent with the data. 
Searches that treat the mixing length as a free parameter are therefore more likely to identify one of over-abundant low $\alpha_{\mathrm{MLT}}$ models. This argues strongly for the calculation of more 3D atmospheres which can be combined with 1D models to remove the sensitivity of the models to this parameter \citep{JA19, 2019MNRAS.tmp.2125S}. 

\subsection{Interpretability and model dependency of results}
\label{sec:IGR}

The results presented in this work are model dependent.
They reflect the methodology of the SPI pipeline and the modelling choices made in the underlying stellar evolution calculations. In previous SPI papers \citep{2016ApJ...830...31B,2017ApJ...839..116A,2019A&A...622A.130B} we have discussed the difficulty in fully propagating the biases and uncertainty from choices in the modelling. Here we briefly comment on the interpretability of our results.

The vast literature dedicated to stellar micro- and macro- physics provides many possibilities by which to construct a model. In this work we have focused on CBM in the form of overshoot. In the case of overshoot we have the option whether or not to include the process in the modelling, we can choose which functional form to adopt, whether to vary the efficiency of overshoot or calibrate it, and whether to assume the process is weak or penetrative. Whilst a detailed analysis of the modelling systematics is beyond the scope of the current study, there are some recent results that focus on this very issue. Extensive tests have been carried out to investigate the minimum level of numerical differences in stellar modelling \citep[Christensen-Dalsgaard, et al. accepted]{2019arXiv191204909S}. Paper III in that series of articles will determine how the modeller, and their choices in input physics (and hence systematics) further compound these differences. 
In their analysis of Procyon, \citet{2014ApJ...787..164G} also demonstrated how the inferred mass differs as a function of assumed overshoot efficiency for both radiative and adiabatic overshoot regions. Differences are non-monotonic and  typically of order a few percent between the adiabatic and radiative grids -- similar to differences from choices in the solar abundance \citep{2011A&A...535A..91D}. Future studies, making use of the  gravity-mode period spacing or acoustic glitches, will help constrain overshoot at the respective convective boundaries.

All asteroseismic investigations of this type have assumed standard stellar physics. However, non-canonical effects such as rotation and magnetic fields may play a crucial role. Any missing processes that impact the structure will be compensated by the degrees of freedom available in the modelling. For example, stars with  $M \gtrapprox 1.3 \Mo$ have different angular momentum and magnetic evolution compared to their lower-mass counterparts. The subsequent impact on the stellar structure will be compensated by the free parameters in the modelling, which include initial mass and overshoot efficiency.

It is not obvious \emph{a priori} how changes in the modelling will shift the predictions for correlated stellar parameters. We have hence been mindful not to over-interpret the results presented in this work. Recall, whilst we place high-credence in our sensitivity to the physically meaningful core sizes,  we have limited our inference to clear outliers (HD~49933 and HD~181906) or large-scale systematic trends (increased scatter for $M \gtrapprox 1.3 \Mo$) for the effective overshoot efficiency. For stars with  $M \gtrapprox 1.3 \Mo$, the shallow convective envelope, short mode lifetimes and large asteroseismic uncertainty contribute to the greater uncertainty in their overshoot predictions.

\section{Conclusions}
\label{sec:conclusions}
In this work we highlighted several cases where asteroseismic modelling favours extreme overshoot efficiencies.
Such values would naively produce extended mixed cores far larger than those determined from calibrations to binaries and stellar clusters.
However, much like the mixing length parameter, CBM results are code dependent even with the same mixing formalism implemented.
In addition to the parameterization of the convective boundary mixing, there are subtleties in the way codes identify the point of neutrality, how they treat different boundaries and how efficiently mixing is applied in the case of small convection zones. 

We implemented CBM with a step overshoot formalism in our models and modified MESA (stellar evolution code) and \SPI (asteroseismic pipeline) to better report on the relevant parameters in order to make our results more generalizeable. Newly constructed \SPI input features demonstrate improved sensitivity to the internal physics of the stars -- in particular to convective boundary mixing.
We analysed main-sequence and subgiant stars in the \Kepler \ and \corot \ fields as well as two radial velocity targets in  $\alpha$~Cen~A and Procyon~A. 
These stars vary in their spectral types, asteroseismic precisions and measured quantities capturing the breadth of data quality we can expect from targets in the \TESS field.
One of the key parameters we calculate is an `effective overshooting' efficiency. It reflects the actual extent of mixing given MESA (like other stellar evolution codes) applies truncations in order to more realistically capture astrophysical fluid behaviour.  We demonstrated that by calculating an `effective overshoot' reasonable efficiencies corresponding to physical core-mass sizes are attained. 
Depending on the evolution code used, this helps to explain the high input overshoot values reported in previous asteroseismic studies.

Two of the targets, F-stars in the \corot \ field measured with low precision, nevertheless displayed anomalous CBM properties compared to their counterparts of similar mass.
We used \SPI and its training grid to determine that an incongruence between the radial and non-radial modes was driving the asteroseismic solution to high overshoot values. 
Due to their short mode lifetimes and large line-widths, mode identifications in F-stars are notoriously difficult. 
Through our improvements to \textsc{SPI}, we have the measures in place to flag such challenging \TESS targets.  

In a forthcoming paper we extend \SPI to greater mass range and to investigate overshoot using the constraints offered from double-line eclipsing binaries. 
Our current analysis did not identify a mass dependence in the inferred overshoot efficiencies.
However, we are limited by large errors in our various overshoot determinations. 
This is partly a result of the degeneracies present in asteroseismic analysis which can be better constrained by binary systems.
The widely varied physics in our training-data and the inherently statistical nature of our methodology provides a novel and complementary analysis for these important calibrators of stellar evolution.

\section*{Acknowledgements}
Part of the research leading to the presented results has received funding from the European Research Council under the European Community's Seventh Framework Programme (FP7/2007-2013) / ERC grant agreement no 338251 (StellarAges).
SB acknowledges NSF grant AST-1514676 and NASA grant NNX16AI09G. 
Funding for the Stellar Astrophysics Centre is provided by The Danish National Research Foundation (Grant agreement no.: DNRF106). 
G.C.A thanks Andreas J{\o}rgensen for interesting discussions about convection as well as the attendees at the Aarhus Challenge modelling workshops for their useful comments that helped improve this manuscript significantly. We thank Jakob Mosumgaard for making some of the BASTA results available for comparison.

%%%%%%%%%%%%%%%%%%%%%%%%%%%%%%%%%%%%%%%%%%%%%%%%%%



\bibliographystyle{mnras}
\bibliography{SPIOS} % if your bibtex file is called example.bib

%\begin{thebibliography}{99}
%\end{thebibliography}

%%%%%%%%%%%%%%%%%%%%%%%%%%%%%%%%%%%%%%%%%%%%%%%%%%

%%%%%%%%%%%%%%%%% APPENDICES %%%%%%%%%%%%%%%%%%%%%

\appendix

\section{Seismic Definitions} 
\label{sec:sdefs}
Oscillation modes are defined in terms of their frequency, $\nu$, and regular spherical harmonic quantum numbers $\ell, m, r$. 
The large-frequency separation is defined as
\begin{equation} 
  \Delta\nu_\ell(n) = \nu_{n,\ell} - \nu_{n-1, \ell}
\end{equation}
and the small-frequency separation defined as
\begin{equation}
  \delta\nu_{\ell,\ell+2} = \nu_{n,\ell} - \nu_{n-1,\ell+2}.
\end{equation}

We utilize the asteroseismic ratios as per \citet{2003AA...411..215R}  to mitigate the impact of the surface term. 
The ratios are defined as: 

\begin{equation}    \label{eqn:LSratio}
  \mathrm{r}_{(\ell,\ell +2)}(n) \equiv \frac{\delta\nu_{(\ell, \ell+2)}(n)}{\Delta\nu_{(1-\ell)}(n+\ell)}.
\end{equation}
\begin{equation} 
  \mathrm{r}_{(\ell, 1-\ell)}(n) \equiv \frac{\mathrm{dd}_{(\ell,1-\ell)}(n)}{\Delta\nu_{(1-\ell)}(n+\ell)} \label{eqn:rnl}
\end{equation}
where the five point separations,  dd$_{(\ell,1-\ell)}(n)$,  are defined as:
\begin{align} 
  \mathrm{dd}_{0,1} \equiv \frac{1}{8} \big[&\nu_0(n-1) - 4\nu_1(n-1) \notag\\
                                 &+6\nu_0(n) - 4\nu_1(n) + \nu_0(n+1)\big]\\ 
  \mathrm{dd}_{1,0} \equiv -\frac{1}{8} \big[&\nu_1(n-1) - 4\nu_0(n) \notag\\
                                &+6\nu_1(n) - 4\nu_0(n+1) + \nu_1(n+1)\big]. \label{eqn:dlast}
\end{align}

We calculate dozens of oscillation frequencies per star with the mode sets available dependent on the internal structure of an individual model. We calculate median values as described in 
\citet{2017ApJ...839..116A} which we denote with angular parentheses. Local values of the ratios, calculated at each (possible) radial order, are also retained for every model.  
We make use of the asteroseismic scaling relations \citep{1986ApJ...306L..37U,1995a&a...293...87k}  as a point of comparison for \SPI. They are defined as

\begin{align} 
%\end{equation}
%\begin{equation}
    \frac{R_\ast}{\text{R}_\odot}
    %\simeq
    &\simeq
    \left(
        \frac{\nu_{\max,\ast}}{\nu_{\max,\odot}}
    \right)
    \left(
        \frac{\Delta\nu_\ast}{\Delta\nu_\odot}
    \right)^{-2}
    \left(
        \frac{T_{\text{eff},\ast}}{T_{\text{eff},\odot}}
    \right)^\frac{1}{2} \label{eq:scalingR}
    \\
    \frac{\rho_\ast}{\rho_\odot}
    %\simeq%
    &\simeq
    \left(
        \frac{\Delta\nu_\ast}{\Delta\nu_\odot}
    \right)^{2}
    \left(
        \frac{T_{\text{eff},\ast}}{T_{\text{eff},\odot}}
    \right) \label{eq:scalingRho}
    \\
    \frac{M_\ast}{\text{M}_\odot}
    %\simeq %
    &\simeq
    \left(
        \frac{\nu_{\max,\ast}}{\nu_{\max,\odot}}
    \right)^3
    \left(
        \frac{\Delta\nu_\ast}{\Delta\nu_\odot}
    \right)^{-4}
    \left(
        \frac{T_{\text{eff},\ast}}{T_{\text{eff},\odot}}
    \right)^\frac{3}{2}. \label{eq:scalingM}
%\end{equation}
%\begin{equation}
%\end{equation}
\end{align}
where  $T_\text{eff}$ is the effective temperature of the star, $\nu_{\max}$ is the frequency of maximum oscillation power, and $\Delta\nu$ is the large frequency separation \citep[for further definitions, see, e.g.,][]{basuchaplin2017}. The quantities subscripted with the solar symbol ($\odot$) correspond to the solar values: ${\nu_{\max,\odot} = 3090\pm30~\mu\text{Hz}}$, ${\Delta\nu_\odot = 135.1\pm 0.1~\mu\text{Hz}}$, and ${T_{\text{eff},\odot} = 5772.0\pm0.8~\text{K}}$ \citep{2011ApJ...743..143H, 2016AJ....152...41P}. 

\section{References in Supplementary Material} 
The supplementary star-by-star analysis of \corot and RV targets makes reference to the following work not cited in the main manuscript:
\citet{2005ApJ...633..424A}, \citet{1996ApJ...473..550C}, \citet{2010A&A...512A..54C}, \cite{2010AN....331..949D}, \citet{2012A&A...543A..96E}, \citet{2017A&A...605A..79G}, \citet{2000AJ....119.2428G}, \citet{2013PNAS..11013267G}, \citet{1998SSRv...85..161G}, \citet{1993oee..conf...15G}, \citet{2014A&A...564A.105H}, \citet{2000A&A...355L..27H}, \citet{2018ApJ...864...99J}, \citet{2010A&A...510A.106K}, \citet{2012A&A...544L..13L}, \citet{2013ApJ...769....7L}, \citet{2009ApJ...699..373M}, \citet{2006AJ....131.1163S} and \citet{2006MmSAI..77..411T}.

\section{Evolutionary Phase Classification}
\label{sec:EPC}
\begin{figure}
    \centering
    \includegraphics[trim={1.9cm 0.5cm 1.3cm 0},clip, width=\columnwidth]{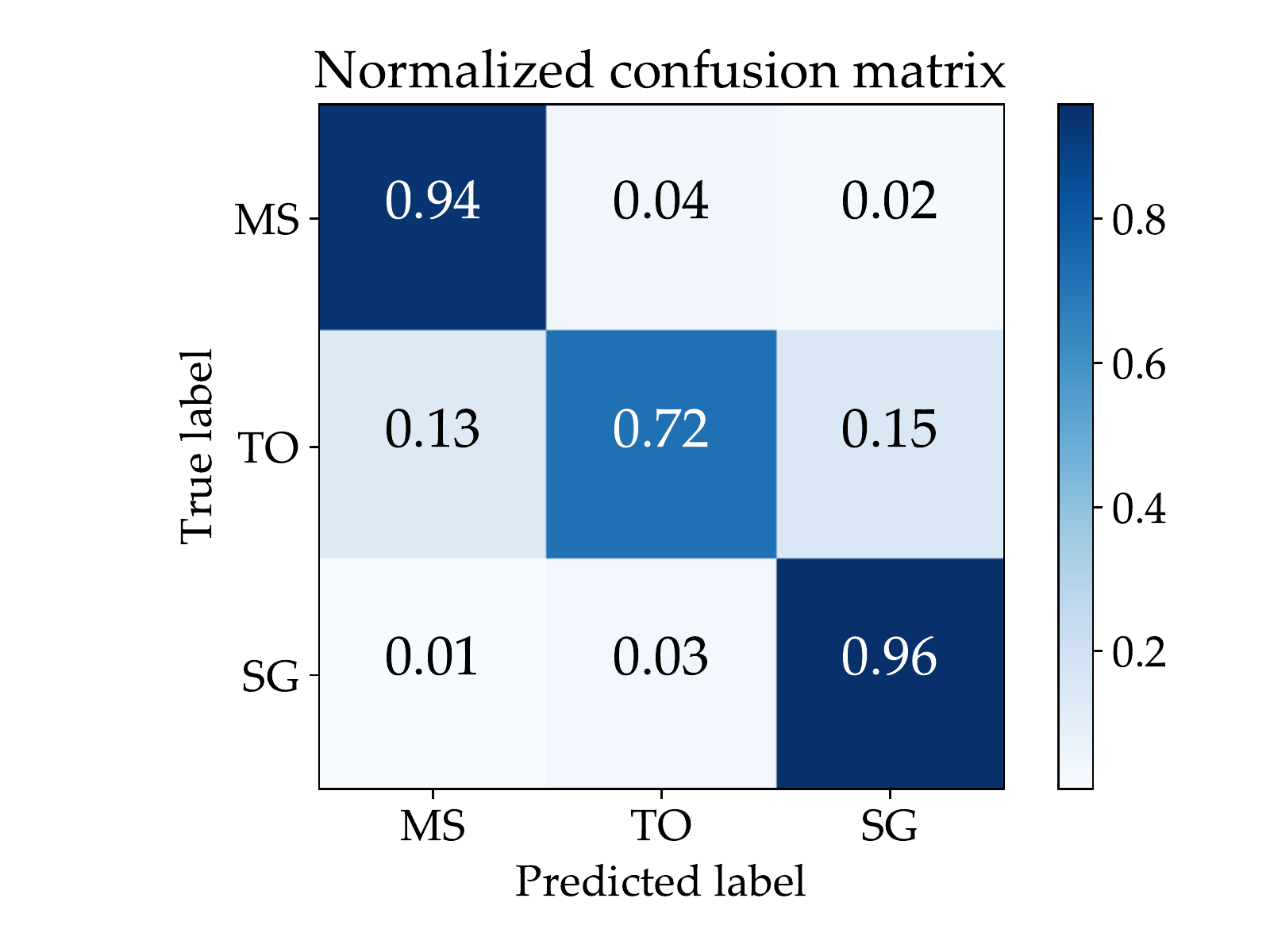}
    \caption{The normalized confusion matrix indicates performance in classifying test data with the respective labels. }
    \label{fig:confusion}
\end{figure}

We have implemented a RF classifier in the \SPI pipeline to determine the evolutionary phase based on  $\nmx$ , $\Teff$, $\Dnu$, and $\Fe$.
Evolutionary phases were divided according to three sampling regimes listed in \S \ref{sec:SPIov}. 
We report a mean accuracy of 91\% based on a withheld development set.  
The confusion matrix (Figure \ref{fig:confusion}) indicates that the classifier can accurately distinguish main-sequence stars from subgiants. This algorithm maximizes the size of the training data and optimally selects input features for each target.

\section{Fit to the fully-mixed mass}
\begin{figure}
    \centering
    \includegraphics[width=\columnwidth]{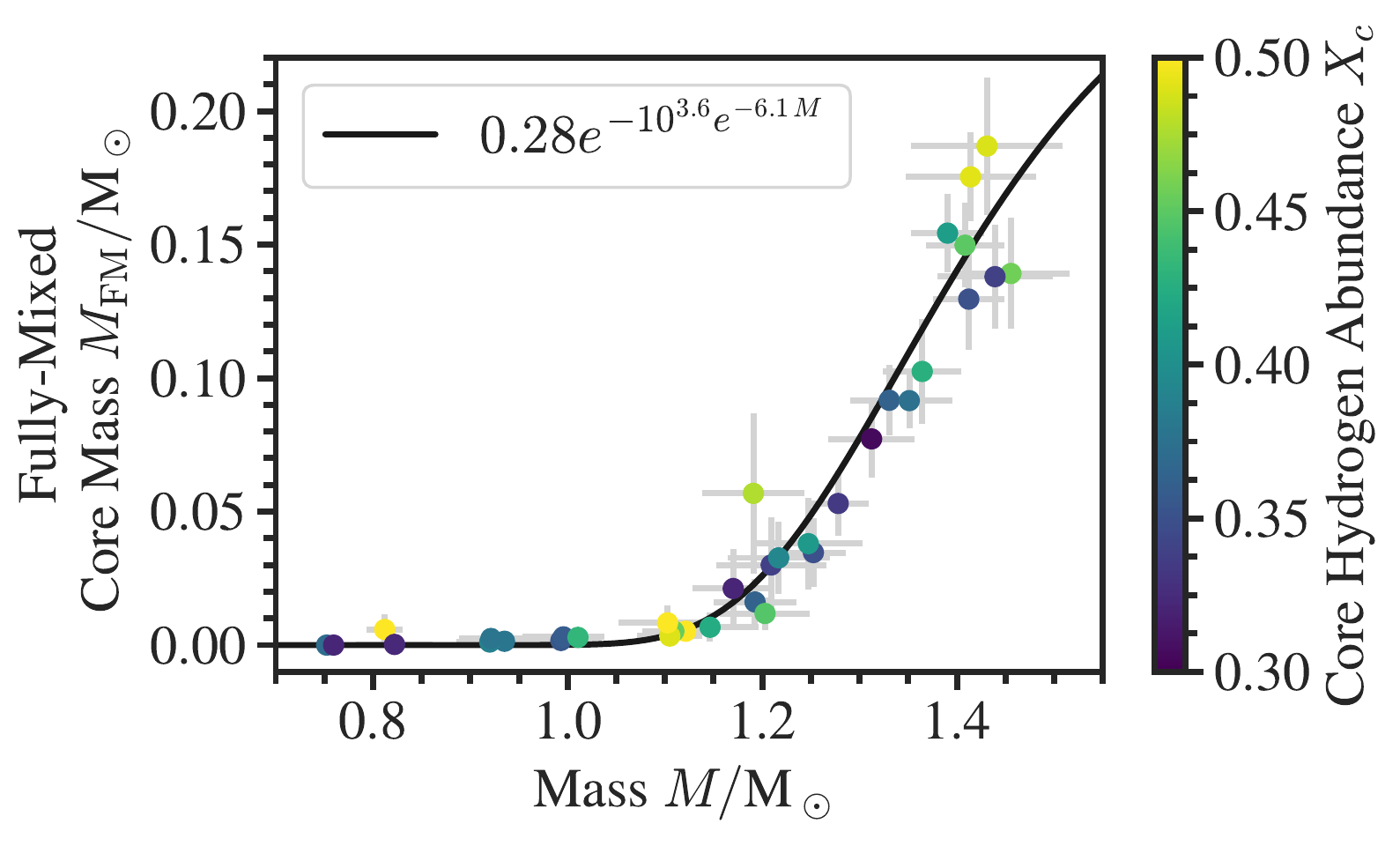}
    \caption{Sigmoidal fit to the fully-mixed core mass for stars with $0.3 < X_c < 0.5$. }
    \label{fig:sigmoid}
\end{figure}

For stars in Figure \ref{fig:OScomparison} with core-hydrogen depletion ($X_c > 0.3$), the fully-mixed core mass can be well-described by a sigmoidal function as described in Figure~\ref{fig:sigmoid}). 

% Don't change these lines
\bsp	% typesetting comment
\label{lastpage}
\end{document}